\documentclass[10pt,prx,aps,twocolumn,showpacs,amsmath,amssymb]{revtex4-1}
\usepackage{times}
\usepackage{graphicx}
\usepackage{psfrag}
\usepackage{gensymb}
\usepackage{epsfig}
\usepackage{color}
\usepackage{afterpage}
\usepackage[tight]{subfigure}
\usepackage{bm}
\allowdisplaybreaks 
\usepackage{braket}
\usepackage[breaklinks]{hyperref}
\hypersetup{colorlinks=true, linkcolor=blue, citecolor=blue, filecolor=blue, urlcolor=blue}

\begin{document}

\title{Universal signatures of Fermi arcs in quasiparticle interference on the surface of Weyl semimetals}

\author{Stefanos Kourtis}
\author{Jian Li}
\author{Zhijun Wang}
\author{Ali Yazdani}
\author{B. Andrei Bernevig}
\affiliation{Department of Physics, Princeton University, Princeton, NJ 08544, USA}

\date{\today}

\begin{abstract}
Weyl semimetals constitute a newly discovered class of three-dimensional topological materials with linear touchings of valence and conduction bands in the bulk. The most striking property of topological origin in these materials, so far unequivocally observed only in photoemission experiments, is the presence of open constant-energy contours at the boundary --- the so-called Fermi arcs. In this work, we establish the universal characteristics of Fermi-arc contributions to surface quasiparticle interference. Using a general phenomenological model, we determine the defining interference patterns stemming from the existence of Fermi arcs in a surface band structure. We then trace these patterns in both simple tight-binding models and realistic ab initio calculations. Our results show that definitive signatures of Fermi arcs can be observed in existing and proposed Weyl semimetals using scanning tunneling spectroscopy.
\end{abstract}

\maketitle

\section{Introduction}

The prototypical examples of noninteracting topological states of matter are categorized by quantized invariants, corresponding to (sets of) energy bands that are separated by gaps from the rest of the band structure. A conceptual step forward in the topological characterization of materials was the definition of topological invariants for systems in which energy gaps vanish and bands touch~\cite{volovikbook}. For example, in the presence of inversion and time-reversal, two-dimensional spinless graphene exhibits a quantized topological invariant.  In the vicinity of isolated points in the Brillouin zone, where Dirac nodes occur, the topological invariant is obtained by integrating the Berry potential over a closed line encircling these points.

Similarly, in three dimensions, nodes may appear in pairs of opposite chirality, i.e., as sources and sinks of Berry flux~\cite{Nielsen1981a,Nielsen1981b,Nielsen1983, volovikbook, Fang2003}. The two nodes in each pair can be pushed apart in reciprocal space by breaking the product of  time reversal and inversion symmetries. The low-energy theory describing electrons at such a nodal point is encapsulated in the Weyl equation. When the chemical potential crosses or is close to these nodal points in a material, the latter is called a Weyl semimetal (WSM)~\cite{Wan2011,Xu2011,Balents2011,Bernevig2015}. Unlike Dirac nodes in graphene, Weyl nodes cannot be gapped or otherwise removed from the band structure by small translation-symmetry preserving perturbations. When a closed Fermi surface (FS) patch encloses only one Weyl node, one can define a FS Chern number, which is equal to the topological charge of the node~\cite{Turner2013,Wang2012,Wang2013a,Weng2014}.

Recently, experimental evidence for the discovery of Weyl fermions in TaAs and NbAs was provided by angle-resolved photoemission spectrocopy~\cite{Lv2015a,Lv2015b,Xu2015a,Xu2015b,Yang2015} and (magneto)transport measurements~\cite{Zhang2015,Zhang2015a}. The theory that guided the discovery~\cite{Weng2015,Huang2015c} attracted immediate attention, because the materials are stoichiometric and therefore easy to synthesize. The prediction of a second type of WSMs rendered another two compounds, WTe${}_2$ and MoTe${}_2$, promising candidates for realization~\cite{Soluyanov2015,Sun2015,Wang2015}.

One of the most interesting hallmarks of a WSM is the presence of open constant-energy contours in its surface band structure called Fermi arcs~\cite{Wan2011,Xu2011}. The existence of the corresponding surface states is a direct consequence of the nonzero topological charge associated with a Weyl node. Since they pertain solely to the surface, these previously elusive FS features are also amenable to observation via scanning tunneling spectroscopy (STS). An analysis of quasiparticle-interference (QPI) patterns in the Fourier-transformed local density of states (FTLDOS) at the boundary of a material can yield important properties of surface quantum states~\cite{McElroy2003,Aynajian2012,Roushan2009,Zhang2009,Seo2010,Okada2011,Alpichshev2010,Alpichshev2011,Fang2013,Zhang2014a}.
The potential for detecting Fermi arcs with STS was recognized in earlier theoretical work~\cite{Hosur2012,Hofmann2013}, but the QPI fingerprints of Fermi arcs remain theoretically and experimentally unresolved.

The purpose of the present manuscript is to determine the unique signatures of Fermi arcs in the QPI patterns obtained by STS measurements at the surface of a WSM. First, we identify the most elementary QPI pattern shapes in the presence of a single Fermi arc and define criteria for their unambiguous experimental observation. Since both discovered and candidate WSMs host two or more pairs of Weyl nodes and will hence have more than one Fermi arcs on a given surface, we examine the fundamental QPI features when more than one arcs coexist on the same surface. In the case of type-2 WSMs, the boundary FS will comprise of both Fermi arcs and electron and hole pockets. We therefore study the fate of the nontrivial characteristics in QPI when surface modes are allowed to scatter into states originating from the bulk. We then pinpoint all aforementioned signatures in QPI patterns obtained from both generic tight-binding models and density functional theory (DFT) calculations for MoTe${}_2$ and TaAs.

\section{Theory of QPI at the surface of Weyl semimetals}

\subsection{Definition of QPI response}

The FTLDOS obtained from STS measurements can be generally expressed as~\cite{Capriotti2003,Derry2015}
\begin{align}
  &F(\bm{q},E) = \frac{i}{2\pi}[\Lambda(\bm{q},E)-\Lambda(-\bm{q},E)^*],  \label{eq:ftldos}\\
  &\Lambda(\bm{q},E) = \int \mathrm{d}\bm{k}\, \text{Tr}[G(\bm{k}+\bm{q},E)T(\bm{k}+\bm{q},\bm{k};E)G(\bm{k},E)] \,,
\end{align}
where $G(\bm{k},E)$ is the retarded Green's function for a clean sample and $T(\bm{k},\bm{k}';E)$ is the $T$-matrix associated with disorder~\cite{Mahan}. On heuristic grounds, the power spectrum $|F(\bm{q},E)|$ is commonly approximated by the autocorrelation of the spectral functions~\cite{Hoffman2002,Simon2007,Roushan2009}
\begin{align}
 &J_{\nu}(\bm{q},E) = \int\mathrm{d}\bm{k} \, \text{Tr}[A_{\nu}(\bm{k}+\bm{q},E) A_{\nu}(\bm{k},E)], \label{eq:J} \\
 &A_{\nu}(\bm{k},E) = (i/2\pi)\text{Tr}_{\bar{\nu}}[G(\bm{k},E)-G(\bm{k},E)^\dag], \label{eq:A}
\end{align}
where $\nu$ stands for the set of inner degrees of freedom that is preserved in the scattering (e.g. spin in spin-preserving scattering), and Tr$_{\bar{\nu}}$ stands for the partial trace over all inner degrees of freedom other than $\nu$, such that $A_{\nu}$ is a reduced density matrix in terms of $\nu$. In this work, we will consider two types of autocorrelations: the joint density of states (JDOS) $J_0$ with $\nu$ being an empty set, and the spin-dependent scattering probability (SSP) $J_s$ with $\nu$ being solely electron spin. The JDOS is particularly important in studying a WSM that lacks any symmetry --- this is the most generic WSM although it is still to be found experimentally; the SSP includes suppressions due to the symmetries of the eigenstates and is hence important for WSMs that respect time-reversal symmetry --- the case for all confirmed WSMs.

The JDOS ignores all matrix-element effects inherent in FTLDOS and takes into account all energetically allowed scattering wavevectors on equal footing, whereas SSP includes only the scattering suppression that comes from the spin content of the wavefunction. Approximating FTLDOS with JDOS / SSP amounts to replacing the impurity landscape with a single scattering center, which can be easily treated within band theory. Even though the rationale behind evaluating JDOS / SSP instead of the full FTLDOS is clear, it is not always straightforward to rigorously connect one to the other~\cite{Derry2015}. For this reason, we have verified that our key findings based on JDOS / SSP calculations are qualitatively the same in the full FTLDOS of our tight-binding models~\cite{suppl}.

\subsection{Phenomenology}

\begin{figure}[t]
 \centering
 \includegraphics[width=\columnwidth]{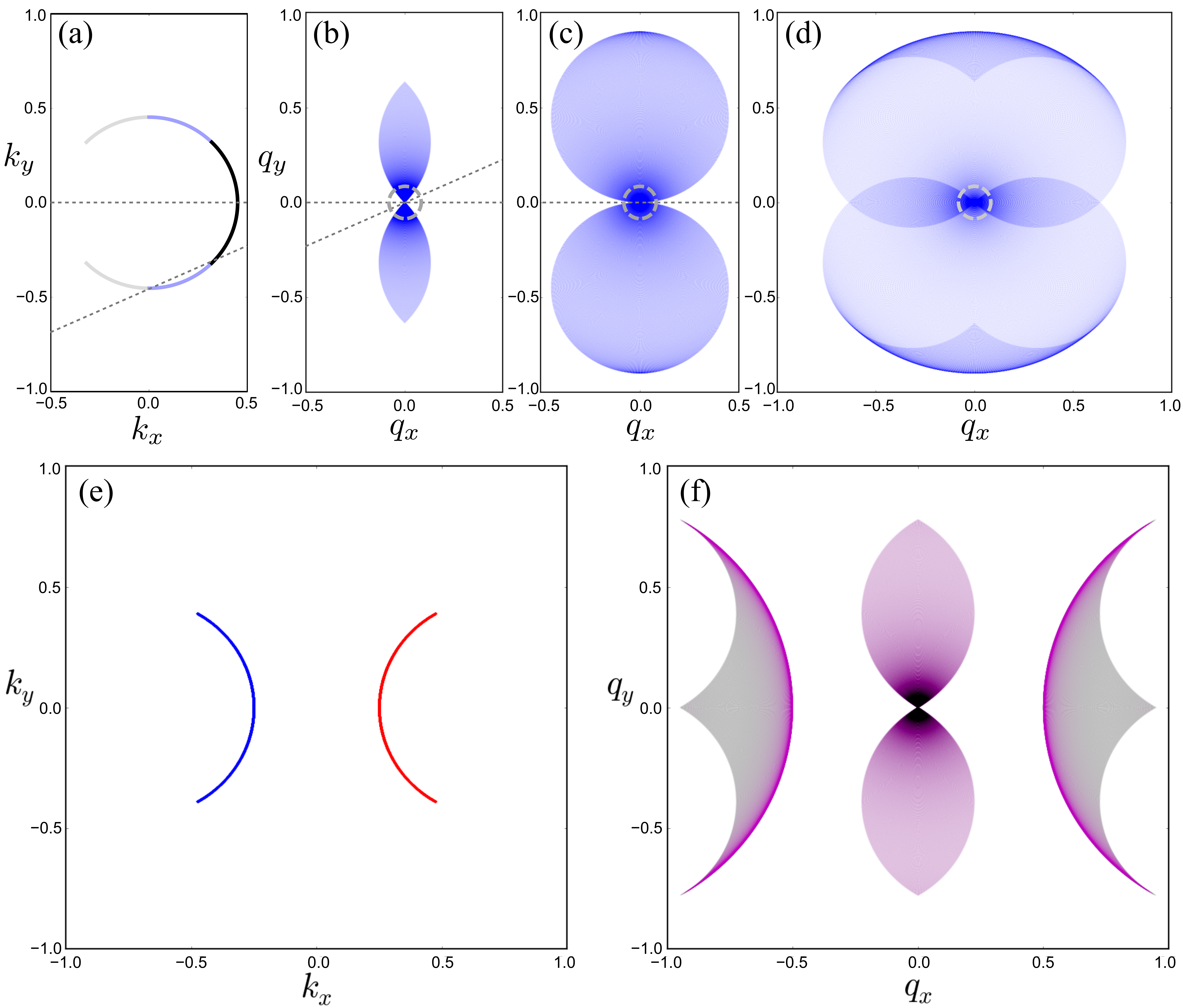}
 \caption{(a-d) Single Fermi arc (a) parametrized by Eq.~\eqref{eq:arc} with $r_1=0.45$, $\bm{k}_1 = 0$, $\gamma_1 = - 2 \varphi_1$ and the shape of its corresponding JDOS from Eq.~\eqref{eq:J} for (b) $\varphi_1 = - \pi/4$, (c) $\varphi_1 = - \pi/2$ and (d) $\varphi_1 = - 3\pi/4$. (e) Two Fermi arcs with $\gamma_1 = \gamma_2 = 2\pi/3$ and (f) the shape of the corresponding JDOS. In (a-d), dashed lines encircle the pinch points; dotted lines are described in the text.}
 \label{fig:toy}
\end{figure}

Let us now consider the JDOS patterns most broadly associated with Fermi arcs. First, we illustrate the key points phenomenologically, by assuming that the Fermi arcs have a constant curvature and a constant spectral density. The Fermi level is supposed to cross the bulk band structure only at the nodal points, so that only boundary modes are visible in the surface FS. The spectral function of an individual arc at a fixed energy can be parametrized as
\begin{equation}
 A(\bm{k};\bm{k}_1,r_1,\gamma_1,\varphi_1) = \int\displaylimits_{\varphi_1}^{\varphi_1+\gamma_1} \mathrm{d}\varphi \, \delta( \bm{k} - \bm{k}_1 - r_1 (\cos\varphi,\sin\varphi) ) \,,\label{eq:arc}
\end{equation}
where $\bm{k}_1$ is the offset of the circle center from the origin, $r_1$ the circle radius and $\gamma_1$ the angle subtended by the arc. The endpoints of the arc are located at $r_1 (\cos\varphi_1,\sin\varphi_1)$ and $r_1 (\cos(\varphi_1+\gamma_1),\sin(\varphi_1+\gamma_1))$. The JDOS generated solely by this single arc is independent of $\bm{k}_1$, while $r_1$ and $\varphi_1$ change only its size and orientation, respectively. The only parameter that affects the shape of the arc is $\gamma_1$. This is shown in Fig.~\ref{fig:toy}(a-d) for three idealized, perfectly circular, cases. Figs.~\ref{fig:toy}(e,f) illustrate the autocorrelation of a FS that includes a second arc. Apart from the feature that arises from the autocorrelations of the two arcs, which is exactly like that of Fig.~\ref{fig:toy}(b), there are now cross-correlation patterns at finite momenta, corresponding to scattering between arcs.

The most distinctive feature is the presence of a pinch point at $\bm{q}=0$ for arcs with $\gamma_1 \leq \pi$. This is a unique characteristic of an open contour in the surface BZ and can be interpreted as follows: a pinch point exists as long as scattering within a FS contour vanishes at all wavevectors along a specific direction. When such a pinch point exists in the QPI pattern, then the contour that generates it must be open. Consider a translation of the spectral function of an arc defined as ${\cal T}_{\epsilon \bm{v}} A(\bm{k}) = A(\bm{k}+\epsilon \bm{v})$, with $\bm{v}$ a unitary vector defining a direction in $\bm{k}$-space and $\epsilon\in \mathbb{R}$. A pinch point exists if there is a $\bm{v}$ such that $A(\bm{k}) {\cal T}_{\epsilon \bm{v}} A(\bm{k}) = 0$ for any $\epsilon\not=0$, so that, from Eq.~\eqref{eq:J}, $J_0(\bm{q}=\epsilon\bm{v})=0$. The directions $\bm{v}$ for which this property holds are revealed by the orientation of the resulting pattern in $J_0$. This is illustrated by the examples in Figs.~\ref{fig:toy}(a-c): a translation of the arc with $\gamma_1 = \pi/2$ [shown in black in Fig.~\ref{fig:toy}(a)] along either of the two dotted lines in Fig.~\ref{fig:toy}(a) leads to $A(\bm{k}) {\cal T}_{\epsilon \bm{v}} A(\bm{k}) = 0$. Translated to the origin, the same lines cross the autocorrelation pattern Fig.~\ref{fig:toy}(b) only at the pinch point. For $\gamma_1 = \pi$, the above holds only for $\bm{v} = \hat{\bm{x}}$, the unitary vector in the $x$ direction. For $\gamma_1 > \pi$, this property does not hold: $A(\bm{k}) {\cal T}_{\epsilon \bm{v}} A(\bm{k}) \not= 0$ for small $\epsilon$ along any $\bm{v}$. Nonetheless, a pinch point can still be found in the autocorrelation of an arc with $\gamma_1 > \pi$: one can simply split it into two arcs, the first one with $\gamma_1'=\pi$ and a second one with the residual angle $\gamma_1-\gamma_1'$. The autocorrelation of the first part generates the pattern in Fig.~\ref{fig:toy}(c), while the autocorrelation of the residue is similar to Fig.~\ref{fig:toy}(b) with a pinch point at $\bm{q}=0$. The pinch point in this case, however, is on top of the pattern stemming from $\gamma_1'$ and the cross-correlation between the two parts [see Fig.~\ref{fig:toy}(d)]. Even though for the purpose of illustration we employed circular arcs, the translation condition for the presence of a pinch point in $J_\nu$ is general and can be used regardless of the arc shape. We shall recover this feature in both tight-binding and DFT calculations below. We remark that, even though the $\bm{q}\simeq0$ region may be difficult to resolve in QPI experiments, identification of the figure-eight pattern at larger $\bm{q}$, like the ones in Figs.~\ref{fig:toy}(b,c,f), indicates a pinch point at $\bm{q}=0$.

\subsection{Tight-binding formulation}

The simplest tight-binding formalism for WSMs is given by the Hamiltonian
\begin{equation}
 {\cal H} = \sum_{\bm{k}} \psi_{\bm{k}}^\dagger H({\bm{k}}) \psi_{\bm{k}} \,,
\end{equation}
where $\psi_{\bm{k}} = (c_{\bm{k},A,\uparrow} \ c_{\bm{k},A,\downarrow} \ c_{\bm{k},B,\uparrow} \ c_{\bm{k},B,\downarrow} )^{\mathsf{T}}$ is a fermionic spinor containing electronic annihilation operators $c_{\bm{k},s,\sigma}$, with $s=A,B$ orbital/sublattice and $\sigma=\uparrow,\downarrow$ spin indices respectively, and $\psi_{\bm{k}}^\dagger$ its hermitian conjugate. Let us first ignore the spin degree of freedom. In this case, we can write a minimal (two-component) tight-binding model describing a WSM with only two Weyl nodes as
\begin{subequations}
\begin{equation}
 H_{2\times2}(\bm{k}) = \bm{g}(\bm{k}) \cdot \boldsymbol\tau + g_{0}(\bm{k}) \tau_0 \,,
\end{equation}
where $\boldsymbol\tau$ is the vector of Pauli matrices and $\tau_0$ the $2\times2$ unity matrix in orbital/sublattice space, $\bm{g} = ( g_{1}, g_{2}, g_{3} )$ and
\begin{align}
 g_{0}(\bm{k}) =&{\ } 2d (2 - \cos k_x - \cos k_y) \,,\\
 g_{1}(\bm{k}) =&{\ } a \sin k_x \,,\\
 g_{2}(\bm{k}) =&{\ } a \sin k_y \,,\\
 g_{3}(\bm{k}) =&{\ } m + t \cos k_z + 2b ( 2 - \cos k_x - \cos k_y ) \,,
\end{align}\label{eq:tbm1}%
\end{subequations}%
with $a, b, d, m, t$ real parameters $(a,t\ne 0)$. With $b=d=0$ and $|m| < |t|$, the energy spectrum has 8 Weyl nodes at points given by $k_{x/y}=0,\pi$ and $k_z = \pm\arccos\frac{m}{t}$. A finite $b$ can gap the nodes with $k_{x/y}=\pi$, so that for $|m+4b|>|t|$ there are exactly two Weyl nodes at $(0,0,\pm\arccos\frac{m}{t})$. If one introduces a boundary, a Fermi arc connects the projections of the nodal points on the boundary FS and $d$ controls the curvature of the arc.

To investigate inter-arc scattering that is subject to time-reversal symmetry, we use the four-spinor $\psi_{\bm{k}}$ and construct the following Hamiltonian
\begin{align}\label{eq:tbm2}
 H_{4\times4}(\bm{k}) =&{\ }g_{1}(\bm{k}) \tau_1 \sigma_3 + g_{2}(\bm{k}) \tau_2 \sigma_0 + g_{3}(\bm{k}) \tau_3 \sigma_0 \nonumber\\
 &{\ } + g_{0}(\bm{k})  \tau_0 \sigma_0 + \beta \tau_2 \sigma_2 + \alpha \sin k_y \tau_1 \sigma_2 \,,
\end{align}
where $\alpha, \beta$ are real parameters, $\sigma_0$ and $\sigma_1, \sigma_2, \sigma_3$ are the $2\times2$ identity and Pauli matrices spanning the spin degree of freedom and a tensor product between $\tau$ and $\sigma$ matrices is assumed. This model produces four Weyl nodes and two Fermi arcs per surface in a finite parameter regime.

\begin{figure}[t]
 \centering
 \includegraphics[width=\columnwidth]{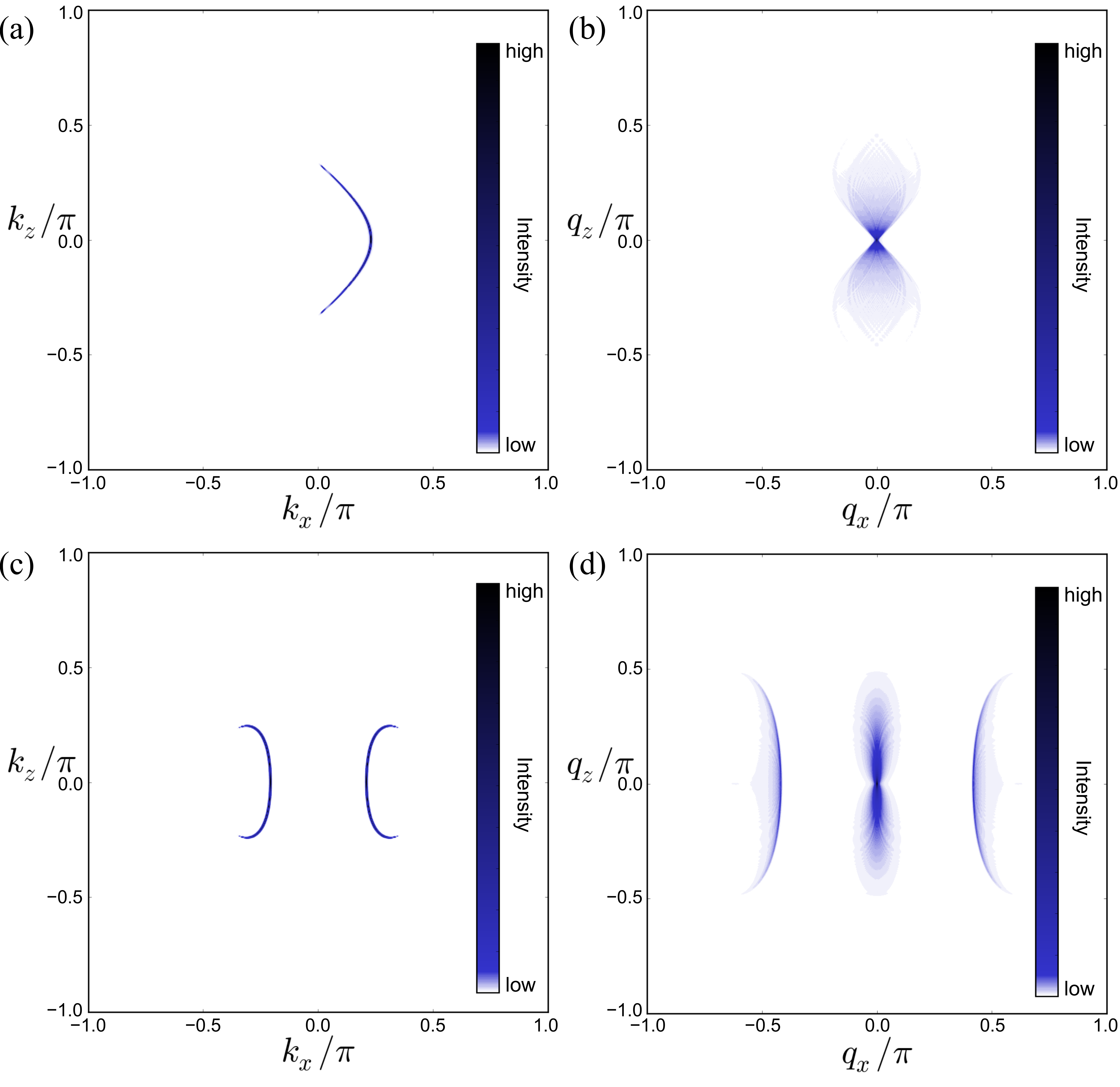}
 \caption{(a,c) Fermi surfaces ($E=0$) projected to the (010) surface; (b) JDOS for (a); (d) SSP for (c). The model for (a) and (b) is Eqs.~\eqref{eq:tbm1} with $a=b=t=1$, $m=0.5$, $d=0.8$; the model for (c) and (d) is Eq.~\eqref{eq:tbm2} with $a=b=1$, $t=1.5$, $d=m=0$, $\beta=0.9$ and $\alpha=0.3$. The JDOS of (c), not shown here, is similar to (d) but shows significantly stronger inter-arc scattering intensity.}
 \label{fig:tbm1}
\end{figure}

Our results for $J_0$ and $J_s$, for one and two Fermi arcs yielded by Eqs.~\eqref{eq:tbm1} and Eqs.~\eqref{eq:tbm2} respectively, are shown in Fig.~\ref{fig:tbm1}~\cite{suppl}. The characteristic ``figure-eight'' encountered in the previous section is evident here as well, but its intensity is modulated in accordance with the Fermi-arc DOS, which causes a fading of the pattern at larger $\bm{q}$. In the case of $H_{4\times4}(\bm{k})$, the suppression due to the spin texture of the two Fermi arcs has been taken into account. As can be seen in the resulting QPI pattern Fig.~\ref{fig:tbm1}(d), there is no qualitative change to the intra-arc scattering intensity, whereas now inter-arc cross-correlation patterns are present [cf. Fig.~\ref{fig:toy}(f)], even though the spin content of the wavefunction causes their partial suppression.

\subsection{Density functional theory}

\begin{figure}[t]
 \centering
 \includegraphics[width=\columnwidth]{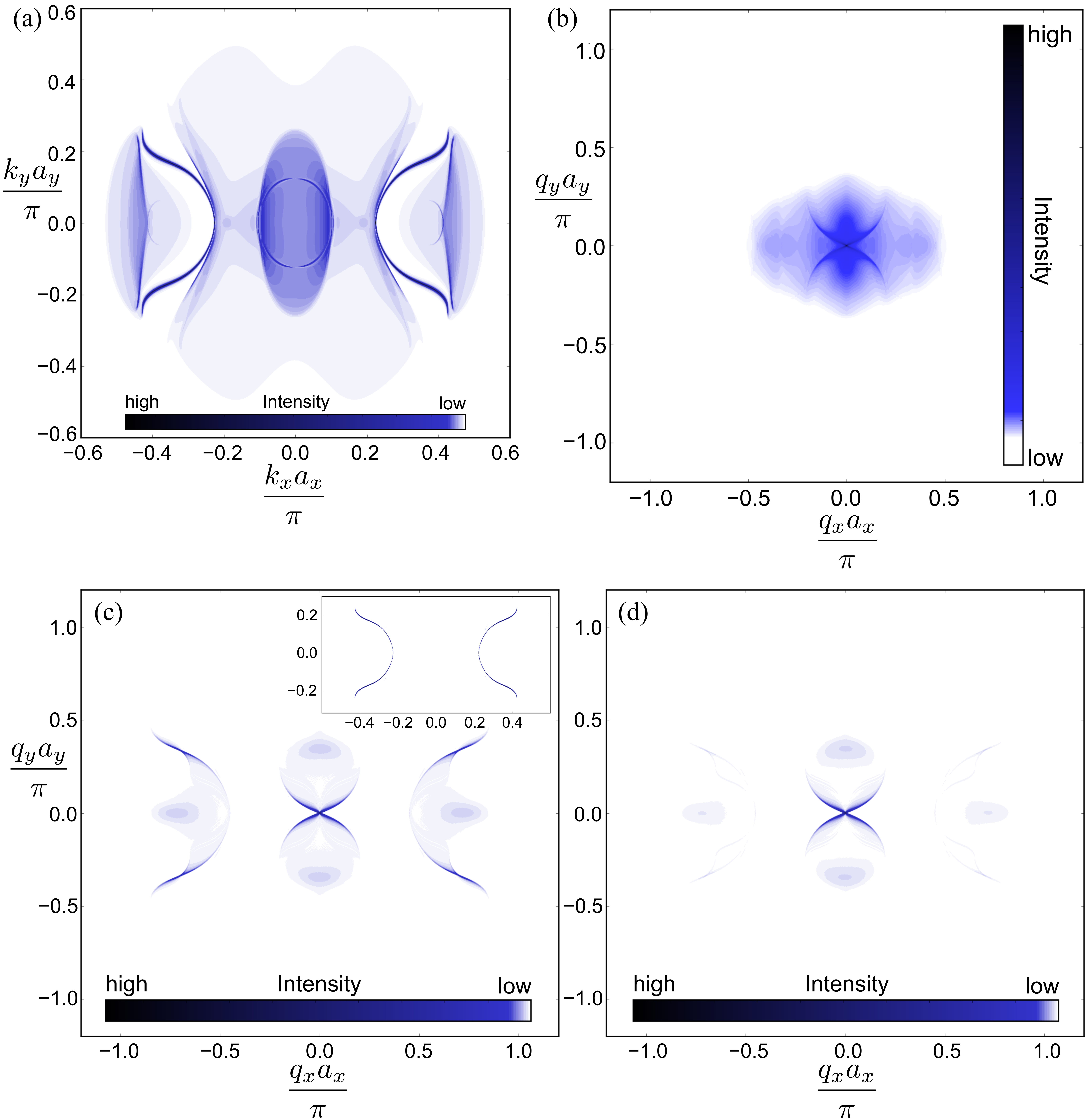}
 \caption{(a) Surface FS and (b) SSP for the (001) surface of MoTe${}_2$ at $E=-0.05$~eV; (c) JDOS for the surface DOS at $\bm{k}$-points where the intensity is not lower than 10\% of the maximum, i.e., keeping only the Fermi arcs shown in the inset of (c); (d) SSP for inset of (c).}
 \label{fig:mote2}
\end{figure}

Finally, we present results for QPI in the experimentally discovered WSMs based on density-functional theory (DFT). First, we focus on MoTe${}_2$, which was recently proposed as a candidate for a type-2 WSM \cite{Sun2015,Wang2015}. The band structure obtained in ab initio calculations features four Weyl nodes at points $(\pm 0.1011 ,\pm 0.0503 , 0)$ in units of reciprocal lattice vectors. This renders the plane $k_y=0$ to be topologically $Z_2$ nontrivial, exhibiting a Quantum Spin Hall effect. The result of this QSH is to give rise to two Fermi arcs per surface. By its definition, a type-2 WSM will have a surface DOS that comprises of both Fermi arcs and bulk states projected to the boundary, which is  shown in Fig.~\ref{fig:mote2}(a). As depicted in Fig.~\ref{fig:mote2}(b), contributions to the JDOS from both types of features are superimposed. Nevertheless, due to the fact that the states of Fermi arcs are more localized on the surface and have a larger intensity compared to the bulk states that participate in the surface DOS, we recover a clear signature of the Fermi arcs in the JDOS in the form of an ``X''-shaped scar. To positively identify this signature, in Fig.~\ref{fig:mote2}(c) we show the JDOS obtained if we ``mask out'' all the bulk signal in the surface DOS. The resulting pattern, which matches the ``X''-shaped feature in Fig.~\ref{fig:mote2}(b) perfectly, is closely resemblant of Figs.~\ref{fig:toy}(f) and~\ref{fig:tbm1}(d). Taking spin suppression into account [see Fig.~\ref{fig:mote2}(d)] does not alter this result significantly: both intra- and inter-arc features are present in the QPI pattern, although the inter-arc part is weaker. This observation shows that it is possible to distill the contribution of Fermi arcs in the surface QPI spectrum, especially for large Fermi arcs, even if the latter comprises of overlapping patterns stemming from arcs and other FS features.

\begin{figure}[t]
 \centering
 \includegraphics[width=\columnwidth]{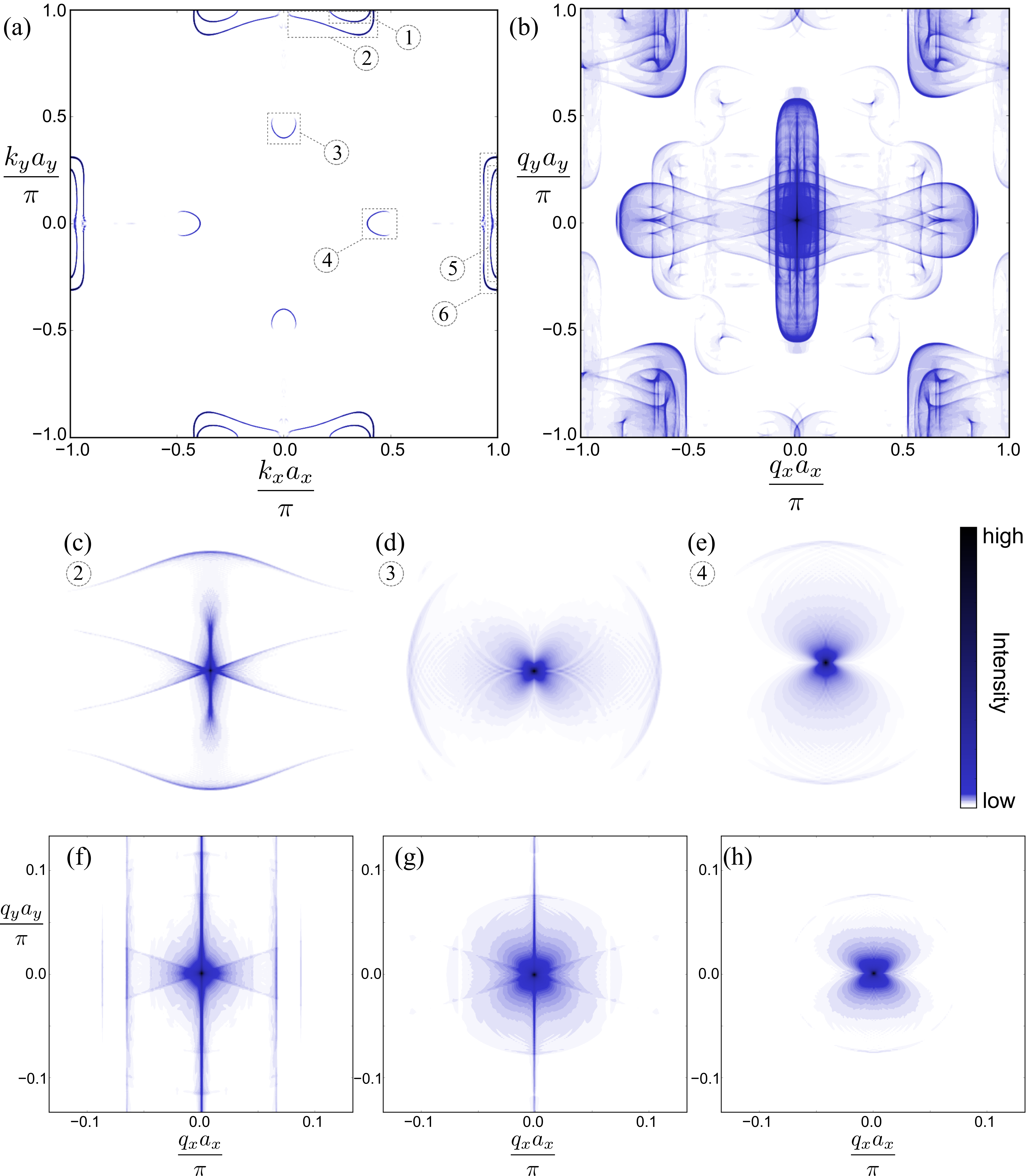}
 \caption{(a) Surface FS and (b) SSP for the (001) surface of TaAs at $E=0.12$~eV; (c-e) autocorrelation of DOS features numbered in (a) --- cf. Figs.~\ref{fig:toy}(b) and~\ref{fig:tbm1}(b); (f) SSP close to $\bm{q}=0$; (g) SSP close to $\bm{q}=0$ minus SSPs centered at $(\pm2\pi,0)$ and $(0,\pm2\pi)$~\cite{suppl}; (h) sum of autocorrelations of features numbered 3 and 4 and their symmetric partners. The intensity of feature 4 is more than two times that of feature 3, so the pattern in (h) is mostly due to the former.}
 \label{fig:taas}
\end{figure}

Next, we investigate the calculated QPI patterns for TaAs. This material has a more complex surface band structure with several Fermi arcs on the (001) surface ~\cite{Weng2015,Huang2015c,Sun2015a}.  The surface DOS obtained from DFT and the corresponding QPI patterns corresponding to the first BZ are presented in Fig.~\ref{fig:taas}. At $E=0.12$~eV bulk contributions to the surface DOS are almost completely suppressed. The FS comprises of 12 Fermi arcs (features 2, 3, 4, 6 and their symmetric copies in Fig.~\ref{fig:taas}) and a smaller number of other, non-topological surface features~\cite{note1}. The bow-tie shaped arcs numbered 2 and 6 extend into the second BZ. With sufficiently high resolution data on a high quality sample all the contributions of the arcs to the QPI should be observable and comparable to our theory. Here, as the $\gamma_1$ angle of the weaker spoon-like features 3 and 4 is less than $\pi$, we focus on identifying the signatures associated with their intra-arc scattering. We can partially isolate their contributions close to $\bm{q}=0$ using only the SSP, as described in the Supplemental Material~\cite{suppl}. With this procedure, we can resolve the figure-eight pattern and pinch point on top of bow-tie contributions, as shown in Figs.~\ref{fig:taas}(g,h). However,  it is likely that the small spoon features observed in our calculation may be obscured by long wavelength variations that typically complicate the analysis of STM QPI data at small $\bm{q}$. 

\section{Conclusion}

In conclusion, we have identified signatures of Fermi arcs in quasiparticle interference at the surface of WSMs. We have observed a characteristic figure-eight shape with a pinch point in its middle in both general tight-binding models and realistic DFT calculations, which, in addition to detailed comparison that can be done for the QPI from the Fermi arcs, is a hallmark of scattering between Fermi arcs. Finally, we have demonstrated that the trademark of a Fermi arc can be distinguished even in cases where the QPI pattern is a superposition of bulk and surface contributions, provided that the Fermi arc has a prominent surface DOS. Our results suggest that there can be an unequivocal observation of Fermi-arc signatures in STS experiments.

\textit{Note} --- Recently, an article with results on quasiparticle interference in Weyl semimetals appeared~\cite{Mitchell2016}. While the results of Ref.~\cite{Mitchell2016} on surface projections of nontrivial bulk topology are complementary to our own, the results on QPI of Fermi arcs show heavy suppression of intra-arc scattering.

\begin{acknowledgments}

This work was supported by NSF CAREER DMR-0952428, ONR-N00014-11-1-0635, MURI-130-6082, the Packard Foundation, and the Keck grant. SK acknowledges financial support by the ICAM branch contributions. JL acknowledges support from Swiss National Science Foundation. SK is grateful to A.~G.~Grushin for enlightening discussions.
\end{acknowledgments}


\begin{thebibliography}{55}%
\makeatletter
\providecommand \@ifxundefined [1]{%
 \@ifx{#1\undefined}
}%
\providecommand \@ifnum [1]{%
 \ifnum #1\expandafter \@firstoftwo
 \else \expandafter \@secondoftwo
 \fi
}%
\providecommand \@ifx [1]{%
 \ifx #1\expandafter \@firstoftwo
 \else \expandafter \@secondoftwo
 \fi
}%
\providecommand \natexlab [1]{#1}%
\providecommand \enquote  [1]{``#1''}%
\providecommand \bibnamefont  [1]{#1}%
\providecommand \bibfnamefont [1]{#1}%
\providecommand \citenamefont [1]{#1}%
\providecommand \href@noop [0]{\@secondoftwo}%
\providecommand \href [0]{\begingroup \@sanitize@url \@href}%
\providecommand \@href[1]{\@@startlink{#1}\@@href}%
\providecommand \@@href[1]{\endgroup#1\@@endlink}%
\providecommand \@sanitize@url [0]{\catcode `\\12\catcode `\$12\catcode
  `\&12\catcode `\#12\catcode `\^12\catcode `\_12\catcode `\%12\relax}%
\providecommand \@@startlink[1]{}%
\providecommand \@@endlink[0]{}%
\providecommand \url  [0]{\begingroup\@sanitize@url \@url }%
\providecommand \@url [1]{\endgroup\@href {#1}{\urlprefix }}%
\providecommand \urlprefix  [0]{URL }%
\providecommand \Eprint [0]{\href }%
\providecommand \doibase [0]{http://dx.doi.org/}%
\providecommand \selectlanguage [0]{\@gobble}%
\providecommand \bibinfo  [0]{\@secondoftwo}%
\providecommand \bibfield  [0]{\@secondoftwo}%
\providecommand \translation [1]{[#1]}%
\providecommand \BibitemOpen [0]{}%
\providecommand \bibitemStop [0]{}%
\providecommand \bibitemNoStop [0]{.\EOS\space}%
\providecommand \EOS [0]{\spacefactor3000\relax}%
\providecommand \BibitemShut  [1]{\csname bibitem#1\endcsname}%
\let\auto@bib@innerbib\@empty
\bibitem [{\citenamefont {Volovik}(2003)}]{volovikbook}%
  \BibitemOpen
  \bibfield  {author} {\bibinfo {author} {\bibfnamefont {G.~E.}\ \bibnamefont
  {Volovik}},\ }\href
  {https://global.oup.com/academic/product/the-universe-in-a-helium-droplet-9780198507826}
  {\emph {\bibinfo {title} {{The Universe in a Helium Droplet}}}}\ (\bibinfo
  {publisher} {Clarendon},\ \bibinfo {address} {Oxford},\ \bibinfo {year}
  {2003})\BibitemShut {NoStop}%
\bibitem [{\citenamefont {Nielsen}\ and\ \citenamefont
  {Ninomiya}(1981{\natexlab{a}})}]{Nielsen1981a}%
  \BibitemOpen
  \bibfield  {author} {\bibinfo {author} {\bibfnamefont {H.~B.}\ \bibnamefont
  {Nielsen}}\ and\ \bibinfo {author} {\bibfnamefont {M.}~\bibnamefont
  {Ninomiya}},\ }\href {\doibase 10.1016/0550-3213(81)90361-8} {\bibfield
  {journal} {\bibinfo  {journal} {Nucl. Phys. B}\ }\textbf {\bibinfo {volume}
  {185}},\ \bibinfo {pages} {20} (\bibinfo {year}
  {1981}{\natexlab{a}})}\BibitemShut {NoStop}%
\bibitem [{\citenamefont {Nielsen}\ and\ \citenamefont
  {Ninomiya}(1981{\natexlab{b}})}]{Nielsen1981b}%
  \BibitemOpen
  \bibfield  {author} {\bibinfo {author} {\bibfnamefont {H.~B.}\ \bibnamefont
  {Nielsen}}\ and\ \bibinfo {author} {\bibfnamefont {M.}~\bibnamefont
  {Ninomiya}},\ }\href {\doibase 10.1016/0550-3213(81)90524-1} {\bibfield
  {journal} {\bibinfo  {journal} {Nucl. Phys. B}\ }\textbf {\bibinfo {volume}
  {193}},\ \bibinfo {pages} {173} (\bibinfo {year}
  {1981}{\natexlab{b}})}\BibitemShut {NoStop}%
\bibitem [{\citenamefont {Nielsen}\ and\ \citenamefont
  {Ninomiya}(1983)}]{Nielsen1983}%
  \BibitemOpen
  \bibfield  {author} {\bibinfo {author} {\bibfnamefont {H.~B.}\ \bibnamefont
  {Nielsen}}\ and\ \bibinfo {author} {\bibfnamefont {M.}~\bibnamefont
  {Ninomiya}},\ }\href {\doibase 10.1016/0370-2693(83)91529-0} {\bibfield
  {journal} {\bibinfo  {journal} {Phys. Lett. B}\ }\textbf {\bibinfo {volume}
  {130}},\ \bibinfo {pages} {389} (\bibinfo {year} {1983})}\BibitemShut
  {NoStop}%
\bibitem [{\citenamefont {Fang}\ \emph {et~al.}(2003)\citenamefont {Fang},
  \citenamefont {Nagaosa}, \citenamefont {Takahashi}, \citenamefont {Asamitsu},
  \citenamefont {Mathieu}, \citenamefont {Ogasawara}, \citenamefont {Yamada},
  \citenamefont {Kawasaki}, \citenamefont {Tokura},\ and\ \citenamefont
  {Terakura}}]{Fang2003}%
  \BibitemOpen
  \bibfield  {author} {\bibinfo {author} {\bibfnamefont {Z.}~\bibnamefont
  {Fang}}, \bibinfo {author} {\bibfnamefont {N.}~\bibnamefont {Nagaosa}},
  \bibinfo {author} {\bibfnamefont {K.~S.}\ \bibnamefont {Takahashi}}, \bibinfo
  {author} {\bibfnamefont {A.}~\bibnamefont {Asamitsu}}, \bibinfo {author}
  {\bibfnamefont {R.}~\bibnamefont {Mathieu}}, \bibinfo {author} {\bibfnamefont
  {T.}~\bibnamefont {Ogasawara}}, \bibinfo {author} {\bibfnamefont
  {H.}~\bibnamefont {Yamada}}, \bibinfo {author} {\bibfnamefont
  {M.}~\bibnamefont {Kawasaki}}, \bibinfo {author} {\bibfnamefont
  {Y.}~\bibnamefont {Tokura}}, \ and\ \bibinfo {author} {\bibfnamefont
  {K.}~\bibnamefont {Terakura}},\ }\href {\doibase 10.1126/science.1089408}
  {\bibfield  {journal} {\bibinfo  {journal} {Science}\ }\textbf {\bibinfo
  {volume} {302}},\ \bibinfo {pages} {92} (\bibinfo {year} {2003})}\BibitemShut
  {NoStop}%
\bibitem [{\citenamefont {Wan}\ \emph {et~al.}(2011)\citenamefont {Wan},
  \citenamefont {Turner}, \citenamefont {Vishwanath},\ and\ \citenamefont
  {Savrasov}}]{Wan2011}%
  \BibitemOpen
  \bibfield  {author} {\bibinfo {author} {\bibfnamefont {X.}~\bibnamefont
  {Wan}}, \bibinfo {author} {\bibfnamefont {A.~M.}\ \bibnamefont {Turner}},
  \bibinfo {author} {\bibfnamefont {A.}~\bibnamefont {Vishwanath}}, \ and\
  \bibinfo {author} {\bibfnamefont {S.~Y.}\ \bibnamefont {Savrasov}},\ }\href
  {\doibase 10.1103/PhysRevB.83.205101} {\bibfield  {journal} {\bibinfo
  {journal} {Phys. Rev. B}\ }\textbf {\bibinfo {volume} {83}},\ \bibinfo
  {pages} {205101} (\bibinfo {year} {2011})}\BibitemShut {NoStop}%
\bibitem [{\citenamefont {Xu}\ \emph {et~al.}(2011)\citenamefont {Xu},
  \citenamefont {Weng}, \citenamefont {Wang}, \citenamefont {Dai},\ and\
  \citenamefont {Fang}}]{Xu2011}%
  \BibitemOpen
  \bibfield  {author} {\bibinfo {author} {\bibfnamefont {G.}~\bibnamefont
  {Xu}}, \bibinfo {author} {\bibfnamefont {H.}~\bibnamefont {Weng}}, \bibinfo
  {author} {\bibfnamefont {Z.}~\bibnamefont {Wang}}, \bibinfo {author}
  {\bibfnamefont {X.}~\bibnamefont {Dai}}, \ and\ \bibinfo {author}
  {\bibfnamefont {Z.}~\bibnamefont {Fang}},\ }\href {\doibase
  10.1103/PhysRevLett.107.186806} {\bibfield  {journal} {\bibinfo  {journal}
  {Phys. Rev. Lett.}\ }\textbf {\bibinfo {volume} {107}},\ \bibinfo {pages}
  {186806} (\bibinfo {year} {2011})}\BibitemShut {NoStop}%
\bibitem [{\citenamefont {Balents}(2011)}]{Balents2011}%
  \BibitemOpen
  \bibfield  {author} {\bibinfo {author} {\bibfnamefont {L.}~\bibnamefont
  {Balents}},\ }\href {\doibase 10.1103/Physics.4.36} {\bibfield  {journal}
  {\bibinfo  {journal} {Physics (College. Park. Md).}\ }\textbf {\bibinfo
  {volume} {4}},\ \bibinfo {pages} {36} (\bibinfo {year} {2011})}\BibitemShut
  {NoStop}%
\bibitem [{\citenamefont {Bernevig}(2015)}]{Bernevig2015}%
  \BibitemOpen
  \bibfield  {author} {\bibinfo {author} {\bibfnamefont {B.~A.}\ \bibnamefont
  {Bernevig}},\ }\href {\doibase 10.1038/nphys3454} {\bibfield  {journal}
  {\bibinfo  {journal} {Nat. Phys.}\ }\textbf {\bibinfo {volume} {11}},\
  \bibinfo {pages} {698} (\bibinfo {year} {2015})}\BibitemShut {NoStop}%
\bibitem [{\citenamefont {Turner}\ and\ \citenamefont
  {Vishwanath}(2013)}]{Turner2013}%
  \BibitemOpen
  \bibfield  {author} {\bibinfo {author} {\bibfnamefont {A.~M.}\ \bibnamefont
  {Turner}}\ and\ \bibinfo {author} {\bibfnamefont {A.}~\bibnamefont
  {Vishwanath}},\ }\href {\doibase 10.1016/B978-0-444-63314-9.00011-1}
  {\bibfield  {journal} {\bibinfo  {journal} {Contemp. Concepts Condens. Matter
  Sci.}\ }\textbf {\bibinfo {volume} {6}},\ \bibinfo {pages} {293} (\bibinfo
  {year} {2013})}\BibitemShut {NoStop}%
\bibitem [{\citenamefont {Wang}\ \emph {et~al.}(2012)\citenamefont {Wang},
  \citenamefont {Sun}, \citenamefont {Chen}, \citenamefont {Franchini},
  \citenamefont {Xu}, \citenamefont {Weng}, \citenamefont {Dai},\ and\
  \citenamefont {Fang}}]{Wang2012}%
  \BibitemOpen
  \bibfield  {author} {\bibinfo {author} {\bibfnamefont {Z.}~\bibnamefont
  {Wang}}, \bibinfo {author} {\bibfnamefont {Y.}~\bibnamefont {Sun}}, \bibinfo
  {author} {\bibfnamefont {X.~Q.}\ \bibnamefont {Chen}}, \bibinfo {author}
  {\bibfnamefont {C.}~\bibnamefont {Franchini}}, \bibinfo {author}
  {\bibfnamefont {G.}~\bibnamefont {Xu}}, \bibinfo {author} {\bibfnamefont
  {H.}~\bibnamefont {Weng}}, \bibinfo {author} {\bibfnamefont {X.}~\bibnamefont
  {Dai}}, \ and\ \bibinfo {author} {\bibfnamefont {Z.}~\bibnamefont {Fang}},\
  }\href {\doibase 10.1103/PhysRevB.85.195320} {\bibfield  {journal} {\bibinfo
  {journal} {Phys. Rev. B}\ }\textbf {\bibinfo {volume} {85}},\ \bibinfo
  {pages} {195320} (\bibinfo {year} {2012})}\BibitemShut {NoStop}%
\bibitem [{\citenamefont {Wang}\ \emph {et~al.}(2013)\citenamefont {Wang},
  \citenamefont {Weng}, \citenamefont {Wu}, \citenamefont {Dai},\ and\
  \citenamefont {Fang}}]{Wang2013a}%
  \BibitemOpen
  \bibfield  {author} {\bibinfo {author} {\bibfnamefont {Z.}~\bibnamefont
  {Wang}}, \bibinfo {author} {\bibfnamefont {H.}~\bibnamefont {Weng}}, \bibinfo
  {author} {\bibfnamefont {Q.}~\bibnamefont {Wu}}, \bibinfo {author}
  {\bibfnamefont {X.}~\bibnamefont {Dai}}, \ and\ \bibinfo {author}
  {\bibfnamefont {Z.}~\bibnamefont {Fang}},\ }\href {\doibase
  10.1103/PhysRevB.88.125427} {\bibfield  {journal} {\bibinfo  {journal} {Phys.
  Rev. B}\ }\textbf {\bibinfo {volume} {88}},\ \bibinfo {pages} {125427}
  (\bibinfo {year} {2013})}\BibitemShut {NoStop}%
\bibitem [{\citenamefont {Weng}\ \emph {et~al.}(2014)\citenamefont {Weng},
  \citenamefont {Dai},\ and\ \citenamefont {Fang}}]{Weng2014}%
  \BibitemOpen
  \bibfield  {author} {\bibinfo {author} {\bibfnamefont {H.}~\bibnamefont
  {Weng}}, \bibinfo {author} {\bibfnamefont {X.}~\bibnamefont {Dai}}, \ and\
  \bibinfo {author} {\bibfnamefont {Z.}~\bibnamefont {Fang}},\ }\href {\doibase
  10.1557/mrs.2014.216} {\bibfield  {journal} {\bibinfo  {journal} {MRS Bull.}\
  }\textbf {\bibinfo {volume} {39}},\ \bibinfo {pages} {849} (\bibinfo {year}
  {2014})}\BibitemShut {NoStop}%
\bibitem [{\citenamefont {Lv}\ \emph {et~al.}(2015{\natexlab{a}})\citenamefont
  {Lv}, \citenamefont {Weng}, \citenamefont {Fu}, \citenamefont {Wang},
  \citenamefont {Miao}, \citenamefont {Ma}, \citenamefont {Richard},
  \citenamefont {Huang}, \citenamefont {Zhao}, \citenamefont {Chen},
  \citenamefont {Fang}, \citenamefont {Dai}, \citenamefont {Qian},\ and\
  \citenamefont {Ding}}]{Lv2015a}%
  \BibitemOpen
  \bibfield  {author} {\bibinfo {author} {\bibfnamefont {B.~Q.}\ \bibnamefont
  {Lv}}, \bibinfo {author} {\bibfnamefont {H.~M.}\ \bibnamefont {Weng}},
  \bibinfo {author} {\bibfnamefont {B.~B.}\ \bibnamefont {Fu}}, \bibinfo
  {author} {\bibfnamefont {X.~P.}\ \bibnamefont {Wang}}, \bibinfo {author}
  {\bibfnamefont {H.}~\bibnamefont {Miao}}, \bibinfo {author} {\bibfnamefont
  {J.}~\bibnamefont {Ma}}, \bibinfo {author} {\bibfnamefont {P.}~\bibnamefont
  {Richard}}, \bibinfo {author} {\bibfnamefont {X.~C.}\ \bibnamefont {Huang}},
  \bibinfo {author} {\bibfnamefont {L.~X.}\ \bibnamefont {Zhao}}, \bibinfo
  {author} {\bibfnamefont {G.~F.}\ \bibnamefont {Chen}}, \bibinfo {author}
  {\bibfnamefont {Z.}~\bibnamefont {Fang}}, \bibinfo {author} {\bibfnamefont
  {X.}~\bibnamefont {Dai}}, \bibinfo {author} {\bibfnamefont {T.}~\bibnamefont
  {Qian}}, \ and\ \bibinfo {author} {\bibfnamefont {H.}~\bibnamefont {Ding}},\
  }\href {\doibase 10.1103/PhysRevX.5.031013} {\bibfield  {journal} {\bibinfo
  {journal} {Phys. Rev. X}\ }\textbf {\bibinfo {volume} {5}},\ \bibinfo {pages}
  {031013} (\bibinfo {year} {2015}{\natexlab{a}})}\BibitemShut {NoStop}%
\bibitem [{\citenamefont {Lv}\ \emph {et~al.}(2015{\natexlab{b}})\citenamefont
  {Lv}, \citenamefont {Xu}, \citenamefont {Weng}, \citenamefont {Ma},
  \citenamefont {Richard}, \citenamefont {Huang}, \citenamefont {Zhao},
  \citenamefont {Chen}, \citenamefont {Matt}, \citenamefont {Bisti},
  \citenamefont {Strocov}, \citenamefont {Mesot}, \citenamefont {Fang},
  \citenamefont {Dai}, \citenamefont {Qian}, \citenamefont {Shi},\ and\
  \citenamefont {Ding}}]{Lv2015b}%
  \BibitemOpen
  \bibfield  {author} {\bibinfo {author} {\bibfnamefont {B.~Q.}\ \bibnamefont
  {Lv}}, \bibinfo {author} {\bibfnamefont {N.}~\bibnamefont {Xu}}, \bibinfo
  {author} {\bibfnamefont {H.~M.}\ \bibnamefont {Weng}}, \bibinfo {author}
  {\bibfnamefont {J.~Z.}\ \bibnamefont {Ma}}, \bibinfo {author} {\bibfnamefont
  {P.}~\bibnamefont {Richard}}, \bibinfo {author} {\bibfnamefont {X.~C.}\
  \bibnamefont {Huang}}, \bibinfo {author} {\bibfnamefont {L.~X.}\ \bibnamefont
  {Zhao}}, \bibinfo {author} {\bibfnamefont {G.~F.}\ \bibnamefont {Chen}},
  \bibinfo {author} {\bibfnamefont {C.~E.}\ \bibnamefont {Matt}}, \bibinfo
  {author} {\bibfnamefont {F.}~\bibnamefont {Bisti}}, \bibinfo {author}
  {\bibfnamefont {V.~N.}\ \bibnamefont {Strocov}}, \bibinfo {author}
  {\bibfnamefont {J.}~\bibnamefont {Mesot}}, \bibinfo {author} {\bibfnamefont
  {Z.}~\bibnamefont {Fang}}, \bibinfo {author} {\bibfnamefont {X.}~\bibnamefont
  {Dai}}, \bibinfo {author} {\bibfnamefont {T.}~\bibnamefont {Qian}}, \bibinfo
  {author} {\bibfnamefont {M.}~\bibnamefont {Shi}}, \ and\ \bibinfo {author}
  {\bibfnamefont {H.}~\bibnamefont {Ding}},\ }\href {\doibase
  10.1038/nphys3426} {\bibfield  {journal} {\bibinfo  {journal} {Nat. Phys.}\
  }\textbf {\bibinfo {volume} {11}},\ \bibinfo {pages} {724} (\bibinfo {year}
  {2015}{\natexlab{b}})}\BibitemShut {NoStop}%
\bibitem [{\citenamefont {Xu}\ \emph {et~al.}(2015{\natexlab{a}})\citenamefont
  {Xu}, \citenamefont {Belopolski}, \citenamefont {Alidoust}, \citenamefont
  {Neupane}, \citenamefont {Bian}, \citenamefont {Zhang}, \citenamefont
  {Sankar}, \citenamefont {Chang}, \citenamefont {Yuan}, \citenamefont {Lee},
  \citenamefont {Huang}, \citenamefont {Zheng}, \citenamefont {Ma},
  \citenamefont {Sanchez}, \citenamefont {Wang}, \citenamefont {Bansil},
  \citenamefont {Chou}, \citenamefont {Shibayev}, \citenamefont {Lin},
  \citenamefont {Jia},\ and\ \citenamefont {Hasan}}]{Xu2015a}%
  \BibitemOpen
  \bibfield  {author} {\bibinfo {author} {\bibfnamefont {S.-Y.}\ \bibnamefont
  {Xu}}, \bibinfo {author} {\bibfnamefont {I.}~\bibnamefont {Belopolski}},
  \bibinfo {author} {\bibfnamefont {N.}~\bibnamefont {Alidoust}}, \bibinfo
  {author} {\bibfnamefont {M.}~\bibnamefont {Neupane}}, \bibinfo {author}
  {\bibfnamefont {G.}~\bibnamefont {Bian}}, \bibinfo {author} {\bibfnamefont
  {C.}~\bibnamefont {Zhang}}, \bibinfo {author} {\bibfnamefont
  {R.}~\bibnamefont {Sankar}}, \bibinfo {author} {\bibfnamefont
  {G.}~\bibnamefont {Chang}}, \bibinfo {author} {\bibfnamefont
  {Z.}~\bibnamefont {Yuan}}, \bibinfo {author} {\bibfnamefont {C.-C.}\
  \bibnamefont {Lee}}, \bibinfo {author} {\bibfnamefont {S.-M.}\ \bibnamefont
  {Huang}}, \bibinfo {author} {\bibfnamefont {H.}~\bibnamefont {Zheng}},
  \bibinfo {author} {\bibfnamefont {J.}~\bibnamefont {Ma}}, \bibinfo {author}
  {\bibfnamefont {D.~S.}\ \bibnamefont {Sanchez}}, \bibinfo {author}
  {\bibfnamefont {B.}~\bibnamefont {Wang}}, \bibinfo {author} {\bibfnamefont
  {A.}~\bibnamefont {Bansil}}, \bibinfo {author} {\bibfnamefont
  {F.}~\bibnamefont {Chou}}, \bibinfo {author} {\bibfnamefont {P.~P.}\
  \bibnamefont {Shibayev}}, \bibinfo {author} {\bibfnamefont {H.}~\bibnamefont
  {Lin}}, \bibinfo {author} {\bibfnamefont {S.}~\bibnamefont {Jia}}, \ and\
  \bibinfo {author} {\bibfnamefont {M.~Z.}\ \bibnamefont {Hasan}},\ }\href
  {\doibase 10.1126/science.aaa9297} {\bibfield  {journal} {\bibinfo  {journal}
  {Science}\ }\textbf {\bibinfo {volume} {349}},\ \bibinfo {pages} {613}
  (\bibinfo {year} {2015}{\natexlab{a}})}\BibitemShut {NoStop}%
\bibitem [{\citenamefont {Xu}\ \emph {et~al.}(2015{\natexlab{b}})\citenamefont
  {Xu}, \citenamefont {Alidoust}, \citenamefont {Belopolski}, \citenamefont
  {Yuan}, \citenamefont {Bian}, \citenamefont {Chang}, \citenamefont {Zheng},
  \citenamefont {Strocov}, \citenamefont {Sanchez}, \citenamefont {Chang},
  \citenamefont {Zhang}, \citenamefont {Mou}, \citenamefont {Wu}, \citenamefont
  {Huang}, \citenamefont {Lee}, \citenamefont {Huang}, \citenamefont {Wang},
  \citenamefont {Bansil}, \citenamefont {Jeng}, \citenamefont {Neupert},
  \citenamefont {Kaminski}, \citenamefont {Lin}, \citenamefont {Jia},\ and\
  \citenamefont {{Zahid Hasan}}}]{Xu2015b}%
  \BibitemOpen
  \bibfield  {author} {\bibinfo {author} {\bibfnamefont {S.-Y.}\ \bibnamefont
  {Xu}}, \bibinfo {author} {\bibfnamefont {N.}~\bibnamefont {Alidoust}},
  \bibinfo {author} {\bibfnamefont {I.}~\bibnamefont {Belopolski}}, \bibinfo
  {author} {\bibfnamefont {Z.}~\bibnamefont {Yuan}}, \bibinfo {author}
  {\bibfnamefont {G.}~\bibnamefont {Bian}}, \bibinfo {author} {\bibfnamefont
  {T.-R.}\ \bibnamefont {Chang}}, \bibinfo {author} {\bibfnamefont
  {H.}~\bibnamefont {Zheng}}, \bibinfo {author} {\bibfnamefont {V.~N.}\
  \bibnamefont {Strocov}}, \bibinfo {author} {\bibfnamefont {D.~S.}\
  \bibnamefont {Sanchez}}, \bibinfo {author} {\bibfnamefont {G.}~\bibnamefont
  {Chang}}, \bibinfo {author} {\bibfnamefont {C.}~\bibnamefont {Zhang}},
  \bibinfo {author} {\bibfnamefont {D.}~\bibnamefont {Mou}}, \bibinfo {author}
  {\bibfnamefont {Y.}~\bibnamefont {Wu}}, \bibinfo {author} {\bibfnamefont
  {L.}~\bibnamefont {Huang}}, \bibinfo {author} {\bibfnamefont {C.-C.}\
  \bibnamefont {Lee}}, \bibinfo {author} {\bibfnamefont {S.-M.}\ \bibnamefont
  {Huang}}, \bibinfo {author} {\bibfnamefont {B.}~\bibnamefont {Wang}},
  \bibinfo {author} {\bibfnamefont {A.}~\bibnamefont {Bansil}}, \bibinfo
  {author} {\bibfnamefont {H.-T.}\ \bibnamefont {Jeng}}, \bibinfo {author}
  {\bibfnamefont {T.}~\bibnamefont {Neupert}}, \bibinfo {author} {\bibfnamefont
  {A.}~\bibnamefont {Kaminski}}, \bibinfo {author} {\bibfnamefont
  {H.}~\bibnamefont {Lin}}, \bibinfo {author} {\bibfnamefont {S.}~\bibnamefont
  {Jia}}, \ and\ \bibinfo {author} {\bibfnamefont {M.}~\bibnamefont {{Zahid
  Hasan}}},\ }\href {\doibase 10.1038/nphys3437} {\bibfield  {journal}
  {\bibinfo  {journal} {Nat. Phys.}\ }\textbf {\bibinfo {volume} {11}},\
  \bibinfo {pages} {748} (\bibinfo {year} {2015}{\natexlab{b}})}\BibitemShut
  {NoStop}%
\bibitem [{\citenamefont {Yang}\ \emph {et~al.}(2015)\citenamefont {Yang},
  \citenamefont {Liu}, \citenamefont {Sun}, \citenamefont {Peng}, \citenamefont
  {Yang}, \citenamefont {Zhang}, \citenamefont {Zhou}, \citenamefont {Zhang},
  \citenamefont {Guo}, \citenamefont {Rahn}, \citenamefont {Prabhakaran},
  \citenamefont {Hussain}, \citenamefont {Mo}, \citenamefont {Felser},
  \citenamefont {Yan},\ and\ \citenamefont {Chen}}]{Yang2015}%
  \BibitemOpen
  \bibfield  {author} {\bibinfo {author} {\bibfnamefont {L.~X.}\ \bibnamefont
  {Yang}}, \bibinfo {author} {\bibfnamefont {Z.~K.}\ \bibnamefont {Liu}},
  \bibinfo {author} {\bibfnamefont {Y.}~\bibnamefont {Sun}}, \bibinfo {author}
  {\bibfnamefont {H.}~\bibnamefont {Peng}}, \bibinfo {author} {\bibfnamefont
  {H.~F.}\ \bibnamefont {Yang}}, \bibinfo {author} {\bibfnamefont
  {T.}~\bibnamefont {Zhang}}, \bibinfo {author} {\bibfnamefont
  {B.}~\bibnamefont {Zhou}}, \bibinfo {author} {\bibfnamefont {Y.}~\bibnamefont
  {Zhang}}, \bibinfo {author} {\bibfnamefont {Y.~F.}\ \bibnamefont {Guo}},
  \bibinfo {author} {\bibfnamefont {M.}~\bibnamefont {Rahn}}, \bibinfo {author}
  {\bibfnamefont {D.}~\bibnamefont {Prabhakaran}}, \bibinfo {author}
  {\bibfnamefont {Z.}~\bibnamefont {Hussain}}, \bibinfo {author} {\bibfnamefont
  {S.-K.}\ \bibnamefont {Mo}}, \bibinfo {author} {\bibfnamefont
  {C.}~\bibnamefont {Felser}}, \bibinfo {author} {\bibfnamefont
  {B.}~\bibnamefont {Yan}}, \ and\ \bibinfo {author} {\bibfnamefont {Y.~L.}\
  \bibnamefont {Chen}},\ }\href {\doibase 10.1038/nphys3425} {\bibfield
  {journal} {\bibinfo  {journal} {Nat. Phys.}\ }\textbf {\bibinfo {volume}
  {11}},\ \bibinfo {pages} {728} (\bibinfo {year} {2015})}\BibitemShut
  {NoStop}%
\bibitem [{\citenamefont {Zhang}\ \emph
  {et~al.}(2015{\natexlab{a}})\citenamefont {Zhang}, \citenamefont {Yuan},
  \citenamefont {Xu}, \citenamefont {Lin}, \citenamefont {Tong}, \citenamefont
  {Hasan}, \citenamefont {Wang}, \citenamefont {Zhang},\ and\ \citenamefont
  {Jia}}]{Zhang2015}%
  \BibitemOpen
  \bibfield  {author} {\bibinfo {author} {\bibfnamefont {C.}~\bibnamefont
  {Zhang}}, \bibinfo {author} {\bibfnamefont {Z.}~\bibnamefont {Yuan}},
  \bibinfo {author} {\bibfnamefont {S.-Y.}\ \bibnamefont {Xu}}, \bibinfo
  {author} {\bibfnamefont {Z.}~\bibnamefont {Lin}}, \bibinfo {author}
  {\bibfnamefont {B.}~\bibnamefont {Tong}}, \bibinfo {author} {\bibfnamefont
  {M.~Z.}\ \bibnamefont {Hasan}}, \bibinfo {author} {\bibfnamefont
  {J.}~\bibnamefont {Wang}}, \bibinfo {author} {\bibfnamefont {C.}~\bibnamefont
  {Zhang}}, \ and\ \bibinfo {author} {\bibfnamefont {S.}~\bibnamefont {Jia}},\
  }\href {http://arxiv.org/abs/1502.00251v1} {\bibfield  {journal} {\bibinfo
  {journal} {arXiv:1502.00251}\ } (\bibinfo {year}
  {2015}{\natexlab{a}})}\BibitemShut {NoStop}%
\bibitem [{\citenamefont {Zhang}\ \emph
  {et~al.}(2015{\natexlab{b}})\citenamefont {Zhang}, \citenamefont {Xu},
  \citenamefont {Belopolski}, \citenamefont {Yuan}, \citenamefont {Lin},
  \citenamefont {Tong}, \citenamefont {Alidoust}, \citenamefont {Lee},
  \citenamefont {Huang}, \citenamefont {Lin}, \citenamefont {Neupane},
  \citenamefont {Sanchez}, \citenamefont {Zheng}, \citenamefont {Bian},
  \citenamefont {Wang}, \citenamefont {Zhang}, \citenamefont {Neupert},
  \citenamefont {Hasan},\ and\ \citenamefont {Jia}}]{Zhang2015a}%
  \BibitemOpen
  \bibfield  {author} {\bibinfo {author} {\bibfnamefont {C.}~\bibnamefont
  {Zhang}}, \bibinfo {author} {\bibfnamefont {S.-Y.}\ \bibnamefont {Xu}},
  \bibinfo {author} {\bibfnamefont {I.}~\bibnamefont {Belopolski}}, \bibinfo
  {author} {\bibfnamefont {Z.}~\bibnamefont {Yuan}}, \bibinfo {author}
  {\bibfnamefont {Z.}~\bibnamefont {Lin}}, \bibinfo {author} {\bibfnamefont
  {B.}~\bibnamefont {Tong}}, \bibinfo {author} {\bibfnamefont {N.}~\bibnamefont
  {Alidoust}}, \bibinfo {author} {\bibfnamefont {C.-C.}\ \bibnamefont {Lee}},
  \bibinfo {author} {\bibfnamefont {S.-M.}\ \bibnamefont {Huang}}, \bibinfo
  {author} {\bibfnamefont {H.}~\bibnamefont {Lin}}, \bibinfo {author}
  {\bibfnamefont {M.}~\bibnamefont {Neupane}}, \bibinfo {author} {\bibfnamefont
  {D.~S.}\ \bibnamefont {Sanchez}}, \bibinfo {author} {\bibfnamefont
  {H.}~\bibnamefont {Zheng}}, \bibinfo {author} {\bibfnamefont
  {G.}~\bibnamefont {Bian}}, \bibinfo {author} {\bibfnamefont {J.}~\bibnamefont
  {Wang}}, \bibinfo {author} {\bibfnamefont {C.}~\bibnamefont {Zhang}},
  \bibinfo {author} {\bibfnamefont {T.}~\bibnamefont {Neupert}}, \bibinfo
  {author} {\bibfnamefont {M.~Z.}\ \bibnamefont {Hasan}}, \ and\ \bibinfo
  {author} {\bibfnamefont {S.}~\bibnamefont {Jia}},\ }\href
  {http://arxiv.org/abs/1503.02630} {\bibfield  {journal} {\bibinfo  {journal}
  {arXiv:1503.02630}\ } (\bibinfo {year} {2015}{\natexlab{b}})}\BibitemShut
  {NoStop}%
\bibitem [{\citenamefont {Weng}\ \emph {et~al.}(2015)\citenamefont {Weng},
  \citenamefont {Fang}, \citenamefont {Fang}, \citenamefont {Bernevig},\ and\
  \citenamefont {Dai}}]{Weng2015}%
  \BibitemOpen
  \bibfield  {author} {\bibinfo {author} {\bibfnamefont {H.}~\bibnamefont
  {Weng}}, \bibinfo {author} {\bibfnamefont {C.}~\bibnamefont {Fang}}, \bibinfo
  {author} {\bibfnamefont {Z.}~\bibnamefont {Fang}}, \bibinfo {author}
  {\bibfnamefont {B.~A.}\ \bibnamefont {Bernevig}}, \ and\ \bibinfo {author}
  {\bibfnamefont {X.}~\bibnamefont {Dai}},\ }\href {\doibase
  10.1103/PhysRevX.5.011029} {\bibfield  {journal} {\bibinfo  {journal} {Phys.
  Rev. X}\ }\textbf {\bibinfo {volume} {5}},\ \bibinfo {pages} {011029}
  (\bibinfo {year} {2015})}\BibitemShut {NoStop}%
\bibitem [{\citenamefont {Huang}\ \emph {et~al.}(2015)\citenamefont {Huang},
  \citenamefont {Xu}, \citenamefont {Belopolski}, \citenamefont {Lee},
  \citenamefont {Chang}, \citenamefont {Wang}, \citenamefont {Alidoust},
  \citenamefont {Bian}, \citenamefont {Neupane}, \citenamefont {Bansil},
  \citenamefont {Lin},\ and\ \citenamefont {Hasan}}]{Huang2015c}%
  \BibitemOpen
  \bibfield  {author} {\bibinfo {author} {\bibfnamefont {S.-M.}\ \bibnamefont
  {Huang}}, \bibinfo {author} {\bibfnamefont {S.-Y.}\ \bibnamefont {Xu}},
  \bibinfo {author} {\bibfnamefont {I.}~\bibnamefont {Belopolski}}, \bibinfo
  {author} {\bibfnamefont {C.-C.}\ \bibnamefont {Lee}}, \bibinfo {author}
  {\bibfnamefont {G.}~\bibnamefont {Chang}}, \bibinfo {author} {\bibfnamefont
  {B.}~\bibnamefont {Wang}}, \bibinfo {author} {\bibfnamefont {N.}~\bibnamefont
  {Alidoust}}, \bibinfo {author} {\bibfnamefont {G.}~\bibnamefont {Bian}},
  \bibinfo {author} {\bibfnamefont {M.}~\bibnamefont {Neupane}}, \bibinfo
  {author} {\bibfnamefont {A.}~\bibnamefont {Bansil}}, \bibinfo {author}
  {\bibfnamefont {H.}~\bibnamefont {Lin}}, \ and\ \bibinfo {author}
  {\bibfnamefont {M.~Z.}\ \bibnamefont {Hasan}},\ }\href {\doibase
  10.1038/ncomms8373} {\bibfield  {journal} {\bibinfo  {journal} {Nat.
  Commun.}\ }\textbf {\bibinfo {volume} {6}},\ \bibinfo {pages} {7373}
  (\bibinfo {year} {2015})}\BibitemShut {NoStop}%
\bibitem [{\citenamefont {Soluyanov}\ \emph {et~al.}(2015)\citenamefont
  {Soluyanov}, \citenamefont {Gresch}, \citenamefont {Wang}, \citenamefont
  {Wu}, \citenamefont {Troyer}, \citenamefont {Dai},\ and\ \citenamefont
  {Bernevig}}]{Soluyanov2015}%
  \BibitemOpen
  \bibfield  {author} {\bibinfo {author} {\bibfnamefont {A.~A.}\ \bibnamefont
  {Soluyanov}}, \bibinfo {author} {\bibfnamefont {D.}~\bibnamefont {Gresch}},
  \bibinfo {author} {\bibfnamefont {Z.}~\bibnamefont {Wang}}, \bibinfo {author}
  {\bibfnamefont {Q.~S.}\ \bibnamefont {Wu}}, \bibinfo {author} {\bibfnamefont
  {M.}~\bibnamefont {Troyer}}, \bibinfo {author} {\bibfnamefont
  {X.}~\bibnamefont {Dai}}, \ and\ \bibinfo {author} {\bibfnamefont {B.~A.}\
  \bibnamefont {Bernevig}},\ }\href {http://arxiv.org/abs/1507.01603}
  {\bibfield  {journal} {\bibinfo  {journal} {arXiv:1507.01603v1}\ } (\bibinfo
  {year} {2015})}\BibitemShut {NoStop}%
\bibitem [{\citenamefont {Sun}\ \emph {et~al.}(2015{\natexlab{a}})\citenamefont
  {Sun}, \citenamefont {Wu}, \citenamefont {Ali}, \citenamefont {Felser},\ and\
  \citenamefont {Yan}}]{Sun2015}%
  \BibitemOpen
  \bibfield  {author} {\bibinfo {author} {\bibfnamefont {Y.}~\bibnamefont
  {Sun}}, \bibinfo {author} {\bibfnamefont {S.-C.}\ \bibnamefont {Wu}},
  \bibinfo {author} {\bibfnamefont {M.~N.}\ \bibnamefont {Ali}}, \bibinfo
  {author} {\bibfnamefont {C.}~\bibnamefont {Felser}}, \ and\ \bibinfo {author}
  {\bibfnamefont {B.}~\bibnamefont {Yan}},\ }\href {\doibase
  10.1103/PhysRevB.92.161107} {\bibfield  {journal} {\bibinfo  {journal} {Phys.
  Rev. B}\ }\textbf {\bibinfo {volume} {92}},\ \bibinfo {pages} {161107}
  (\bibinfo {year} {2015}{\natexlab{a}})}\BibitemShut {NoStop}%
\bibitem [{\citenamefont {Wang}\ \emph {et~al.}(2015)\citenamefont {Wang},
  \citenamefont {Gresch}, \citenamefont {Soluyanov}, \citenamefont {Xie},
  \citenamefont {Dai}, \citenamefont {Troyer}, \citenamefont {Cava},\ and\
  \citenamefont {Bernevig}}]{Wang2015}%
  \BibitemOpen
  \bibfield  {author} {\bibinfo {author} {\bibfnamefont {Z.}~\bibnamefont
  {Wang}}, \bibinfo {author} {\bibfnamefont {D.}~\bibnamefont {Gresch}},
  \bibinfo {author} {\bibfnamefont {A.~A.}\ \bibnamefont {Soluyanov}}, \bibinfo
  {author} {\bibfnamefont {W.}~\bibnamefont {Xie}}, \bibinfo {author}
  {\bibfnamefont {X.}~\bibnamefont {Dai}}, \bibinfo {author} {\bibfnamefont
  {M.}~\bibnamefont {Troyer}}, \bibinfo {author} {\bibfnamefont {R.~J.}\
  \bibnamefont {Cava}}, \ and\ \bibinfo {author} {\bibfnamefont {B.~A.}\
  \bibnamefont {Bernevig}},\ }\href {http://arxiv.org/abs/1511.07440}
  {\bibfield  {journal} {\bibinfo  {journal} {arXiv:1511.07440}\ } (\bibinfo
  {year} {2015})}\BibitemShut {NoStop}%
\bibitem [{\citenamefont {McElroy}\ \emph {et~al.}(2003)\citenamefont
  {McElroy}, \citenamefont {Simmonds}, \citenamefont {Hoffman}, \citenamefont
  {Lee}, \citenamefont {Orenstein}, \citenamefont {Eisaki}, \citenamefont
  {Uchida},\ and\ \citenamefont {Davis}}]{McElroy2003}%
  \BibitemOpen
  \bibfield  {author} {\bibinfo {author} {\bibfnamefont {K.}~\bibnamefont
  {McElroy}}, \bibinfo {author} {\bibfnamefont {R.~W.}\ \bibnamefont
  {Simmonds}}, \bibinfo {author} {\bibfnamefont {J.~E.}\ \bibnamefont
  {Hoffman}}, \bibinfo {author} {\bibfnamefont {D.-H.}\ \bibnamefont {Lee}},
  \bibinfo {author} {\bibfnamefont {J.}~\bibnamefont {Orenstein}}, \bibinfo
  {author} {\bibfnamefont {H.}~\bibnamefont {Eisaki}}, \bibinfo {author}
  {\bibfnamefont {S.}~\bibnamefont {Uchida}}, \ and\ \bibinfo {author}
  {\bibfnamefont {J.~C.}\ \bibnamefont {Davis}},\ }\href {\doibase
  10.1038/nature01496} {\bibfield  {journal} {\bibinfo  {journal} {Nature}\
  }\textbf {\bibinfo {volume} {422}},\ \bibinfo {pages} {592} (\bibinfo {year}
  {2003})}\BibitemShut {NoStop}%
\bibitem [{\citenamefont {Aynajian}\ \emph {et~al.}(2012)\citenamefont
  {Aynajian}, \citenamefont {{da Silva Neto}}, \citenamefont {Gyenis},
  \citenamefont {Baumbach}, \citenamefont {Thompson}, \citenamefont {Fisk},
  \citenamefont {Bauer},\ and\ \citenamefont {Yazdani}}]{Aynajian2012}%
  \BibitemOpen
  \bibfield  {author} {\bibinfo {author} {\bibfnamefont {P.}~\bibnamefont
  {Aynajian}}, \bibinfo {author} {\bibfnamefont {E.~H.}\ \bibnamefont {{da
  Silva Neto}}}, \bibinfo {author} {\bibfnamefont {A.}~\bibnamefont {Gyenis}},
  \bibinfo {author} {\bibfnamefont {R.~E.}\ \bibnamefont {Baumbach}}, \bibinfo
  {author} {\bibfnamefont {J.~D.}\ \bibnamefont {Thompson}}, \bibinfo {author}
  {\bibfnamefont {Z.}~\bibnamefont {Fisk}}, \bibinfo {author} {\bibfnamefont
  {E.~D.}\ \bibnamefont {Bauer}}, \ and\ \bibinfo {author} {\bibfnamefont
  {A.}~\bibnamefont {Yazdani}},\ }\href {\doibase 10.1038/nature11204}
  {\bibfield  {journal} {\bibinfo  {journal} {Nature}\ }\textbf {\bibinfo
  {volume} {486}},\ \bibinfo {pages} {201} (\bibinfo {year}
  {2012})}\BibitemShut {NoStop}%
\bibitem [{\citenamefont {Roushan}\ \emph {et~al.}(2009)\citenamefont
  {Roushan}, \citenamefont {Seo}, \citenamefont {Parker}, \citenamefont {Hor},
  \citenamefont {Hsieh}, \citenamefont {Qian}, \citenamefont {Richardella},
  \citenamefont {Hasan}, \citenamefont {Cava},\ and\ \citenamefont
  {Yazdani}}]{Roushan2009}%
  \BibitemOpen
  \bibfield  {author} {\bibinfo {author} {\bibfnamefont {P.}~\bibnamefont
  {Roushan}}, \bibinfo {author} {\bibfnamefont {J.}~\bibnamefont {Seo}},
  \bibinfo {author} {\bibfnamefont {C.~V.}\ \bibnamefont {Parker}}, \bibinfo
  {author} {\bibfnamefont {Y.~S.}\ \bibnamefont {Hor}}, \bibinfo {author}
  {\bibfnamefont {D.}~\bibnamefont {Hsieh}}, \bibinfo {author} {\bibfnamefont
  {D.}~\bibnamefont {Qian}}, \bibinfo {author} {\bibfnamefont {A.}~\bibnamefont
  {Richardella}}, \bibinfo {author} {\bibfnamefont {M.~Z.}\ \bibnamefont
  {Hasan}}, \bibinfo {author} {\bibfnamefont {R.~J.}\ \bibnamefont {Cava}}, \
  and\ \bibinfo {author} {\bibfnamefont {A.}~\bibnamefont {Yazdani}},\ }\href
  {\doibase 10.1038/nature08308} {\bibfield  {journal} {\bibinfo  {journal}
  {Nature}\ }\textbf {\bibinfo {volume} {460}},\ \bibinfo {pages} {1106}
  (\bibinfo {year} {2009})}\BibitemShut {NoStop}%
\bibitem [{\citenamefont {Zhang}\ \emph {et~al.}(2009)\citenamefont {Zhang},
  \citenamefont {Cheng}, \citenamefont {Chen}, \citenamefont {Jia},
  \citenamefont {Ma}, \citenamefont {He}, \citenamefont {Wang}, \citenamefont
  {Zhang}, \citenamefont {Dai}, \citenamefont {Fang}, \citenamefont {Xie},\
  and\ \citenamefont {Xue}}]{Zhang2009}%
  \BibitemOpen
  \bibfield  {author} {\bibinfo {author} {\bibfnamefont {T.}~\bibnamefont
  {Zhang}}, \bibinfo {author} {\bibfnamefont {P.}~\bibnamefont {Cheng}},
  \bibinfo {author} {\bibfnamefont {X.}~\bibnamefont {Chen}}, \bibinfo {author}
  {\bibfnamefont {J.~F.}\ \bibnamefont {Jia}}, \bibinfo {author} {\bibfnamefont
  {X.}~\bibnamefont {Ma}}, \bibinfo {author} {\bibfnamefont {K.}~\bibnamefont
  {He}}, \bibinfo {author} {\bibfnamefont {L.}~\bibnamefont {Wang}}, \bibinfo
  {author} {\bibfnamefont {H.}~\bibnamefont {Zhang}}, \bibinfo {author}
  {\bibfnamefont {X.}~\bibnamefont {Dai}}, \bibinfo {author} {\bibfnamefont
  {Z.}~\bibnamefont {Fang}}, \bibinfo {author} {\bibfnamefont {X.}~\bibnamefont
  {Xie}}, \ and\ \bibinfo {author} {\bibfnamefont {Q.~K.}\ \bibnamefont
  {Xue}},\ }\href {\doibase 10.1103/PhysRevLett.103.266803} {\bibfield
  {journal} {\bibinfo  {journal} {Phys. Rev. Lett.}\ }\textbf {\bibinfo
  {volume} {103}},\ \bibinfo {pages} {266803} (\bibinfo {year}
  {2009})}\BibitemShut {NoStop}%
\bibitem [{\citenamefont {Seo}\ \emph {et~al.}(2010)\citenamefont {Seo},
  \citenamefont {Roushan}, \citenamefont {Beidenkopf}, \citenamefont {Hor},
  \citenamefont {Cava},\ and\ \citenamefont {Yazdani}}]{Seo2010}%
  \BibitemOpen
  \bibfield  {author} {\bibinfo {author} {\bibfnamefont {J.}~\bibnamefont
  {Seo}}, \bibinfo {author} {\bibfnamefont {P.}~\bibnamefont {Roushan}},
  \bibinfo {author} {\bibfnamefont {H.}~\bibnamefont {Beidenkopf}}, \bibinfo
  {author} {\bibfnamefont {Y.~S.}\ \bibnamefont {Hor}}, \bibinfo {author}
  {\bibfnamefont {R.~J.}\ \bibnamefont {Cava}}, \ and\ \bibinfo {author}
  {\bibfnamefont {A.}~\bibnamefont {Yazdani}},\ }\href {\doibase
  10.1038/nature09189} {\bibfield  {journal} {\bibinfo  {journal} {Nature}\
  }\textbf {\bibinfo {volume} {466}},\ \bibinfo {pages} {343} (\bibinfo {year}
  {2010})}\BibitemShut {NoStop}%
\bibitem [{\citenamefont {Okada}\ \emph {et~al.}(2011)\citenamefont {Okada},
  \citenamefont {Dhital}, \citenamefont {Zhou}, \citenamefont {Huemiller},
  \citenamefont {Lin}, \citenamefont {Basak}, \citenamefont {Bansil},
  \citenamefont {Huang}, \citenamefont {Ding}, \citenamefont {Wang},
  \citenamefont {Wilson},\ and\ \citenamefont {Madhavan}}]{Okada2011}%
  \BibitemOpen
  \bibfield  {author} {\bibinfo {author} {\bibfnamefont {Y.}~\bibnamefont
  {Okada}}, \bibinfo {author} {\bibfnamefont {C.}~\bibnamefont {Dhital}},
  \bibinfo {author} {\bibfnamefont {W.}~\bibnamefont {Zhou}}, \bibinfo {author}
  {\bibfnamefont {E.~D.}\ \bibnamefont {Huemiller}}, \bibinfo {author}
  {\bibfnamefont {H.}~\bibnamefont {Lin}}, \bibinfo {author} {\bibfnamefont
  {S.}~\bibnamefont {Basak}}, \bibinfo {author} {\bibfnamefont
  {A.}~\bibnamefont {Bansil}}, \bibinfo {author} {\bibfnamefont {Y.~B.}\
  \bibnamefont {Huang}}, \bibinfo {author} {\bibfnamefont {H.}~\bibnamefont
  {Ding}}, \bibinfo {author} {\bibfnamefont {Z.}~\bibnamefont {Wang}}, \bibinfo
  {author} {\bibfnamefont {S.~D.}\ \bibnamefont {Wilson}}, \ and\ \bibinfo
  {author} {\bibfnamefont {V.}~\bibnamefont {Madhavan}},\ }\href {\doibase
  10.1103/PhysRevLett.106.206805} {\bibfield  {journal} {\bibinfo  {journal}
  {Phys. Rev. Lett.}\ }\textbf {\bibinfo {volume} {106}},\ \bibinfo {pages}
  {206805} (\bibinfo {year} {2011})}\BibitemShut {NoStop}%
\bibitem [{\citenamefont {Alpichshev}\ \emph {et~al.}(2010)\citenamefont
  {Alpichshev}, \citenamefont {Analytis}, \citenamefont {Chu}, \citenamefont
  {Fisher}, \citenamefont {Chen}, \citenamefont {Shen}, \citenamefont {Fang},\
  and\ \citenamefont {Kapitulnik}}]{Alpichshev2010}%
  \BibitemOpen
  \bibfield  {author} {\bibinfo {author} {\bibfnamefont {Z.}~\bibnamefont
  {Alpichshev}}, \bibinfo {author} {\bibfnamefont {J.~G.}\ \bibnamefont
  {Analytis}}, \bibinfo {author} {\bibfnamefont {J.~H.}\ \bibnamefont {Chu}},
  \bibinfo {author} {\bibfnamefont {I.~R.}\ \bibnamefont {Fisher}}, \bibinfo
  {author} {\bibfnamefont {Y.~L.}\ \bibnamefont {Chen}}, \bibinfo {author}
  {\bibfnamefont {Z.~X.}\ \bibnamefont {Shen}}, \bibinfo {author}
  {\bibfnamefont {A.}~\bibnamefont {Fang}}, \ and\ \bibinfo {author}
  {\bibfnamefont {A.}~\bibnamefont {Kapitulnik}},\ }\href {\doibase
  10.1103/PhysRevLett.104.016401} {\bibfield  {journal} {\bibinfo  {journal}
  {Phys. Rev. Lett.}\ }\textbf {\bibinfo {volume} {104}},\ \bibinfo {pages}
  {016401} (\bibinfo {year} {2010})}\BibitemShut {NoStop}%
\bibitem [{\citenamefont {Alpichshev}\ \emph {et~al.}(2011)\citenamefont
  {Alpichshev}, \citenamefont {Analytis}, \citenamefont {Chu}, \citenamefont
  {Fisher},\ and\ \citenamefont {Kapitulnik}}]{Alpichshev2011}%
  \BibitemOpen
  \bibfield  {author} {\bibinfo {author} {\bibfnamefont {Z.}~\bibnamefont
  {Alpichshev}}, \bibinfo {author} {\bibfnamefont {J.~G.}\ \bibnamefont
  {Analytis}}, \bibinfo {author} {\bibfnamefont {J.~H.}\ \bibnamefont {Chu}},
  \bibinfo {author} {\bibfnamefont {I.~R.}\ \bibnamefont {Fisher}}, \ and\
  \bibinfo {author} {\bibfnamefont {A.}~\bibnamefont {Kapitulnik}},\ }\href
  {\doibase 10.1103/PhysRevB.84.041104} {\bibfield  {journal} {\bibinfo
  {journal} {Phys. Rev. B}\ }\textbf {\bibinfo {volume} {84}},\ \bibinfo
  {pages} {041104} (\bibinfo {year} {2011})}\BibitemShut {NoStop}%
\bibitem [{\citenamefont {Fang}\ \emph {et~al.}(2013)\citenamefont {Fang},
  \citenamefont {Gilbert}, \citenamefont {Xu}, \citenamefont {Bernevig},\ and\
  \citenamefont {Hasan}}]{Fang2013}%
  \BibitemOpen
  \bibfield  {author} {\bibinfo {author} {\bibfnamefont {C.}~\bibnamefont
  {Fang}}, \bibinfo {author} {\bibfnamefont {M.~J.}\ \bibnamefont {Gilbert}},
  \bibinfo {author} {\bibfnamefont {S.-Y.}\ \bibnamefont {Xu}}, \bibinfo
  {author} {\bibfnamefont {B.~A.}\ \bibnamefont {Bernevig}}, \ and\ \bibinfo
  {author} {\bibfnamefont {M.~Z.}\ \bibnamefont {Hasan}},\ }\href {\doibase
  10.1103/PhysRevB.88.125141} {\bibfield  {journal} {\bibinfo  {journal} {Phys.
  Rev. B}\ }\textbf {\bibinfo {volume} {88}},\ \bibinfo {pages} {125141}
  (\bibinfo {year} {2013})}\BibitemShut {NoStop}%
\bibitem [{\citenamefont {Zhang}\ \emph {et~al.}(2014)\citenamefont {Zhang},
  \citenamefont {Baek}, \citenamefont {Ha}, \citenamefont {Zhang},
  \citenamefont {Wyrick}, \citenamefont {Davydov}, \citenamefont {Kuk},\ and\
  \citenamefont {Stroscio}}]{Zhang2014a}%
  \BibitemOpen
  \bibfield  {author} {\bibinfo {author} {\bibfnamefont {D.}~\bibnamefont
  {Zhang}}, \bibinfo {author} {\bibfnamefont {H.}~\bibnamefont {Baek}},
  \bibinfo {author} {\bibfnamefont {J.}~\bibnamefont {Ha}}, \bibinfo {author}
  {\bibfnamefont {T.}~\bibnamefont {Zhang}}, \bibinfo {author} {\bibfnamefont
  {J.}~\bibnamefont {Wyrick}}, \bibinfo {author} {\bibfnamefont {A.~V.}\
  \bibnamefont {Davydov}}, \bibinfo {author} {\bibfnamefont {Y.}~\bibnamefont
  {Kuk}}, \ and\ \bibinfo {author} {\bibfnamefont {J.~A.}\ \bibnamefont
  {Stroscio}},\ }\href {\doibase 10.1103/PhysRevB.89.245445} {\bibfield
  {journal} {\bibinfo  {journal} {Phys. Rev. B}\ }\textbf {\bibinfo {volume}
  {89}},\ \bibinfo {pages} {245445} (\bibinfo {year} {2014})}\BibitemShut
  {NoStop}%
\bibitem [{\citenamefont {Hosur}(2012)}]{Hosur2012}%
  \BibitemOpen
  \bibfield  {author} {\bibinfo {author} {\bibfnamefont {P.}~\bibnamefont
  {Hosur}},\ }\href {\doibase 10.1103/PhysRevB.86.195102} {\bibfield  {journal}
  {\bibinfo  {journal} {Phys. Rev. B}\ }\textbf {\bibinfo {volume} {86}},\
  \bibinfo {pages} {195102} (\bibinfo {year} {2012})}\BibitemShut {NoStop}%
\bibitem [{\citenamefont {Hofmann}\ \emph {et~al.}(2013)\citenamefont
  {Hofmann}, \citenamefont {Queiroz},\ and\ \citenamefont
  {Schnyder}}]{Hofmann2013}%
  \BibitemOpen
  \bibfield  {author} {\bibinfo {author} {\bibfnamefont {J.~S.}\ \bibnamefont
  {Hofmann}}, \bibinfo {author} {\bibfnamefont {R.}~\bibnamefont {Queiroz}}, \
  and\ \bibinfo {author} {\bibfnamefont {A.~P.}\ \bibnamefont {Schnyder}},\
  }\href {\doibase 10.1103/PhysRevB.88.134505} {\bibfield  {journal} {\bibinfo
  {journal} {Phys. Rev. B}\ }\textbf {\bibinfo {volume} {88}},\ \bibinfo
  {pages} {134505} (\bibinfo {year} {2013})}\BibitemShut {NoStop}%
\bibitem [{\citenamefont {Capriotti}\ \emph {et~al.}(2003)\citenamefont
  {Capriotti}, \citenamefont {Scalapino},\ and\ \citenamefont
  {Sedgewick}}]{Capriotti2003}%
  \BibitemOpen
  \bibfield  {author} {\bibinfo {author} {\bibfnamefont {L.}~\bibnamefont
  {Capriotti}}, \bibinfo {author} {\bibfnamefont {D.~J.}\ \bibnamefont
  {Scalapino}}, \ and\ \bibinfo {author} {\bibfnamefont {R.~D.}\ \bibnamefont
  {Sedgewick}},\ }\href {\doibase 10.1103/PhysRevB.68.014508} {\bibfield
  {journal} {\bibinfo  {journal} {Phys. Rev. B}\ }\textbf {\bibinfo {volume}
  {68}},\ \bibinfo {pages} {014508} (\bibinfo {year} {2003})}\BibitemShut
  {NoStop}%
\bibitem [{\citenamefont {Derry}\ \emph {et~al.}(2015)\citenamefont {Derry},
  \citenamefont {Mitchell},\ and\ \citenamefont {Logan}}]{Derry2015}%
  \BibitemOpen
  \bibfield  {author} {\bibinfo {author} {\bibfnamefont {P.~G.}\ \bibnamefont
  {Derry}}, \bibinfo {author} {\bibfnamefont {A.~K.}\ \bibnamefont {Mitchell}},
  \ and\ \bibinfo {author} {\bibfnamefont {D.~E.}\ \bibnamefont {Logan}},\
  }\href {\doibase 10.1103/PhysRevB.92.035126} {\bibfield  {journal} {\bibinfo
  {journal} {Phys. Rev. B}\ }\textbf {\bibinfo {volume} {92}},\ \bibinfo
  {pages} {035126} (\bibinfo {year} {2015})}\BibitemShut {NoStop}%
\bibitem [{\citenamefont {Mahan}(2000)}]{Mahan}%
  \BibitemOpen
  \bibfield  {author} {\bibinfo {author} {\bibfnamefont {G.~D.}\ \bibnamefont
  {Mahan}},\ }\href@noop {} {\emph {\bibinfo {title} {{Many-Particle
  Physics}}}},\ Physics of Solids and Liquids\ (\bibinfo  {publisher}
  {Springer},\ \bibinfo {year} {2000})\BibitemShut {NoStop}%
\bibitem [{\citenamefont {Hoffman}\ \emph {et~al.}(2002)\citenamefont
  {Hoffman}, \citenamefont {McElroy}, \citenamefont {Lee}, \citenamefont
  {Lang}, \citenamefont {Eisaki}, \citenamefont {Uchida},\ and\ \citenamefont
  {Davis}}]{Hoffman2002}%
  \BibitemOpen
  \bibfield  {author} {\bibinfo {author} {\bibfnamefont {J.~E.}\ \bibnamefont
  {Hoffman}}, \bibinfo {author} {\bibfnamefont {K.}~\bibnamefont {McElroy}},
  \bibinfo {author} {\bibfnamefont {D.-H.}\ \bibnamefont {Lee}}, \bibinfo
  {author} {\bibfnamefont {K.~M.}\ \bibnamefont {Lang}}, \bibinfo {author}
  {\bibfnamefont {H.}~\bibnamefont {Eisaki}}, \bibinfo {author} {\bibfnamefont
  {S.}~\bibnamefont {Uchida}}, \ and\ \bibinfo {author} {\bibfnamefont {J.~C.}\
  \bibnamefont {Davis}},\ }\href {\doibase 10.1126/science.1072640} {\bibfield
  {journal} {\bibinfo  {journal} {Science}\ }\textbf {\bibinfo {volume}
  {297}},\ \bibinfo {pages} {1148} (\bibinfo {year} {2002})}\BibitemShut
  {NoStop}%
\bibitem [{\citenamefont {Simon}\ \emph {et~al.}(2007)\citenamefont {Simon},
  \citenamefont {Vonau},\ and\ \citenamefont {Aubel}}]{Simon2007}%
  \BibitemOpen
  \bibfield  {author} {\bibinfo {author} {\bibfnamefont {L.}~\bibnamefont
  {Simon}}, \bibinfo {author} {\bibfnamefont {F.}~\bibnamefont {Vonau}}, \ and\
  \bibinfo {author} {\bibfnamefont {D.}~\bibnamefont {Aubel}},\ }\href
  {\doibase 10.1088/0953-8984/19/35/355009} {\bibfield  {journal} {\bibinfo
  {journal} {J. Phys. Condens. Matter}\ }\textbf {\bibinfo {volume} {19}},\
  \bibinfo {pages} {355009} (\bibinfo {year} {2007})}\BibitemShut {NoStop}%
 \bibitem [{sup()}]{suppl}%
  \BibitemOpen
  \href@noop {} {}\bibinfo {note} {See Supplemental Material for more information on calculations}\BibitemShut {NoStop}%
\bibitem [{\citenamefont {Sun}\ \emph {et~al.}(2015{\natexlab{b}})\citenamefont
  {Sun}, \citenamefont {Wu},\ and\ \citenamefont {Yan}}]{Sun2015a}%
  \BibitemOpen
  \bibfield  {author} {\bibinfo {author} {\bibfnamefont {Y.}~\bibnamefont
  {Sun}}, \bibinfo {author} {\bibfnamefont {S.-C.}\ \bibnamefont {Wu}}, \ and\
  \bibinfo {author} {\bibfnamefont {B.}~\bibnamefont {Yan}},\ }\href {\doibase
  10.1103/PhysRevB.92.115428} {\bibfield  {journal} {\bibinfo  {journal} {Phys.
  Rev. B}\ }\textbf {\bibinfo {volume} {92}},\ \bibinfo {pages} {115428}
  (\bibinfo {year} {2015}{\natexlab{b}})}\BibitemShut {NoStop}%
\bibitem [{sup()}]{note1}%
  \BibitemOpen
  \href@noop {} {}\bibinfo {note} {There are 4 more arcs, located inside features 
  3 and 4 of Fig.~4(a), that are not resolved at this energy}\BibitemShut {NoStop}%
\bibitem [{\citenamefont {Mitchell}\ and\ \citenamefont
  {Fritz}(2016)}]{Mitchell2016}%
  \BibitemOpen
  \bibfield  {author} {\bibinfo {author} {\bibfnamefont {A.~K.}\ \bibnamefont
  {Mitchell}}\ and\ \bibinfo {author} {\bibfnamefont {L.}~\bibnamefont
  {Fritz}},\ }\href {\doibase 10.1103/PhysRevB.93.035137}
  {\bibfield  {journal} {\bibinfo  {journal} {Phys. Rev. B}\ }\textbf {\bibinfo {volume} {93}},\ \bibinfo {pages} {035137} (\bibinfo
  {year} {2016})}\BibitemShut {NoStop}%
\end{thebibliography}

\begin{thebibliography}{3}%
\makeatletter
\providecommand \@ifxundefined [1]{%
 \@ifx{#1\undefined}
}%
\providecommand \@ifnum [1]{%
 \ifnum #1\expandafter \@firstoftwo
 \else \expandafter \@secondoftwo
 \fi
}%
\providecommand \@ifx [1]{%
 \ifx #1\expandafter \@firstoftwo
 \else \expandafter \@secondoftwo
 \fi
}%
\providecommand \natexlab [1]{#1}%
\providecommand \enquote  [1]{``#1''}%
\providecommand \bibnamefont  [1]{#1}%
\providecommand \bibfnamefont [1]{#1}%
\providecommand \citenamefont [1]{#1}%
\providecommand \href@noop [0]{\@secondoftwo}%
\providecommand \href [0]{\begingroup \@sanitize@url \@href}%
\providecommand \@href[1]{\@@startlink{#1}\@@href}%
\providecommand \@@href[1]{\endgroup#1\@@endlink}%
\providecommand \@sanitize@url [0]{\catcode `\\12\catcode `\$12\catcode
  `\&12\catcode `\#12\catcode `\^12\catcode `\_12\catcode `\%12\relax}%
\providecommand \@@startlink[1]{}%
\providecommand \@@endlink[0]{}%
\providecommand \url  [0]{\begingroup\@sanitize@url \@url }%
\providecommand \@url [1]{\endgroup\@href {#1}{\urlprefix }}%
\providecommand \urlprefix  [0]{URL }%
\providecommand \Eprint [0]{\href }%
\providecommand \doibase [0]{http://dx.doi.org/}%
\providecommand \selectlanguage [0]{\@gobble}%
\providecommand \bibinfo  [0]{\@secondoftwo}%
\providecommand \bibfield  [0]{\@secondoftwo}%
\providecommand \translation [1]{[#1]}%
\providecommand \BibitemOpen [0]{}%
\providecommand \bibitemStop [0]{}%
\providecommand \bibitemNoStop [0]{.\EOS\space}%
\providecommand \EOS [0]{\spacefactor3000\relax}%
\providecommand \BibitemShut  [1]{\csname bibitem#1\endcsname}%
\let\auto@bib@innerbib\@empty
\bibitem [{\citenamefont {{L{\'{o}}pez Sancho}}\ \emph
  {et~al.}(1984)\citenamefont {{L{\'{o}}pez Sancho}}, \citenamefont
  {{L{\'{o}}pez Sancho}},\ and\ \citenamefont {Rubio}}]{Sancho1984s}%
  \BibitemOpen
  \bibfield  {author} {\bibinfo {author} {\bibfnamefont {M.~P.}\ \bibnamefont
  {{L{\'{o}}pez Sancho}}}, \bibinfo {author} {\bibfnamefont {J.~M.}\
  \bibnamefont {{L{\'{o}}pez Sancho}}}, \ and\ \bibinfo {author} {\bibfnamefont
  {J.}~\bibnamefont {Rubio}},\ }\href {\doibase 10.1088/0305-4608/14/5/016}
  {\bibfield  {journal} {\bibinfo  {journal} {J. Phys. F Met. Phys.}\ }\textbf
  {\bibinfo {volume} {14}},\ \bibinfo {pages} {1205} (\bibinfo {year}
  {1984})}\BibitemShut {NoStop}%
\bibitem [{\citenamefont {{L{\'{o}}pez Sancho}}\ \emph
  {et~al.}(1985)\citenamefont {{L{\'{o}}pez Sancho}}, \citenamefont
  {{L{\'{o}}pez Sancho}},\ and\ \citenamefont {Rubio}}]{Sancho1985s}%
  \BibitemOpen
  \bibfield  {author} {\bibinfo {author} {\bibfnamefont {M.~P.}\ \bibnamefont
  {{L{\'{o}}pez Sancho}}}, \bibinfo {author} {\bibfnamefont {J.~M.}\
  \bibnamefont {{L{\'{o}}pez Sancho}}}, \ and\ \bibinfo {author} {\bibfnamefont
  {J.}~\bibnamefont {Rubio}},\ }\href {\doibase 10.1088/0305-4608/15/4/009}
  {\bibfield  {journal} {\bibinfo  {journal} {J. Phys. F Met. Phys.}\ }\textbf
  {\bibinfo {volume} {15}},\ \bibinfo {pages} {851} (\bibinfo {year}
  {1985})}\BibitemShut {NoStop}%
\bibitem [{\citenamefont {Dell'Anna}\ \emph {et~al.}(2005)\citenamefont
  {Dell'Anna}, \citenamefont {Lorenzana}, \citenamefont {Capone}, \citenamefont
  {Castellani},\ and\ \citenamefont {Grilli}}]{DellAnna2004}%
  \BibitemOpen
  \bibfield  {author} {\bibinfo {author} {\bibfnamefont {L.}~\bibnamefont
  {Dell'Anna}}, \bibinfo {author} {\bibfnamefont {J.}~\bibnamefont
  {Lorenzana}}, \bibinfo {author} {\bibfnamefont {M.}~\bibnamefont {Capone}},
  \bibinfo {author} {\bibfnamefont {C.}~\bibnamefont {Castellani}}, \ and\
  \bibinfo {author} {\bibfnamefont {M.}~\bibnamefont {Grilli}},\ }\href
  {\doibase 10.1103/PhysRevB.71.064518} {\bibfield  {journal} {\bibinfo
  {journal} {Phys. Rev. B}\ }\textbf {\bibinfo {volume} {71}},\ \bibinfo
  {pages} {064518} (\bibinfo {year} {2005})}\BibitemShut {NoStop}%
\bibitem [{\citenamefont {Choubey}\ \emph {et~al.}(2014)\citenamefont
  {Choubey}, \citenamefont {Berlijn}, \citenamefont {Kreisel}, \citenamefont
  {Cao},\ and\ \citenamefont {Hirschfeld}}]{Choubey2014}%
  \BibitemOpen
  \bibfield  {author} {\bibinfo {author} {\bibfnamefont {P.}~\bibnamefont
  {Choubey}}, \bibinfo {author} {\bibfnamefont {T.}~\bibnamefont {Berlijn}},
  \bibinfo {author} {\bibfnamefont {A.}~\bibnamefont {Kreisel}}, \bibinfo
  {author} {\bibfnamefont {C.}~\bibnamefont {Cao}}, \ and\ \bibinfo {author}
  {\bibfnamefont {P.~J.}\ \bibnamefont {Hirschfeld}},\ }\href {\doibase
  10.1103/PhysRevB.90.134520} {\bibfield  {journal} {\bibinfo  {journal} {Phys.
  Rev. B}\ }\textbf {\bibinfo {volume} {90}},\ \bibinfo {pages} {134520}
  (\bibinfo {year} {2014})}\BibitemShut {NoStop}%
\bibitem [{\citenamefont {Kresse}\ and\ \citenamefont
  {Furthm{\"{u}}ller}(1996)}]{Kresse1996}%
  \BibitemOpen
  \bibfield  {author} {\bibinfo {author} {\bibfnamefont {G.}~\bibnamefont
  {Kresse}}\ and\ \bibinfo {author} {\bibfnamefont {J.}~\bibnamefont
  {Furthm{\"{u}}ller}},\ }\href {\doibase 10.1103/PhysRevB.54.11169} {\bibfield
   {journal} {\bibinfo  {journal} {Phys. Rev. B}\ }\textbf {\bibinfo {volume}
  {54}},\ \bibinfo {pages} {11169} (\bibinfo {year} {1996})}\BibitemShut
  {NoStop}%
\bibitem [{\citenamefont {Bl{\"{o}}chl}(1994)}]{Blochl1994}%
  \BibitemOpen
  \bibfield  {author} {\bibinfo {author} {\bibfnamefont {P.~E.}\ \bibnamefont
  {Bl{\"{o}}chl}},\ }\href {\doibase 10.1103/PhysRevB.50.17953} {\bibfield
  {journal} {\bibinfo  {journal} {Phys. Rev. B}\ }\textbf {\bibinfo {volume}
  {50}},\ \bibinfo {pages} {17953} (\bibinfo {year} {1994})}\BibitemShut
  {NoStop}%
\bibitem [{\citenamefont {Perdew}\ \emph {et~al.}(1996)\citenamefont {Perdew},
  \citenamefont {Burke},\ and\ \citenamefont {Ernzerhof}}]{Perdew1996}%
  \BibitemOpen
  \bibfield  {author} {\bibinfo {author} {\bibfnamefont {J.~P.}\ \bibnamefont
  {Perdew}}, \bibinfo {author} {\bibfnamefont {K.}~\bibnamefont {Burke}}, \
  and\ \bibinfo {author} {\bibfnamefont {M.}~\bibnamefont {Ernzerhof}},\ }\href
  {\doibase 10.1103/PhysRevLett.77.3865} {\bibfield  {journal} {\bibinfo
  {journal} {Phys. Rev. Lett.}\ }\textbf {\bibinfo {volume} {77}},\ \bibinfo
  {pages} {3865} (\bibinfo {year} {1996})}\BibitemShut {NoStop}%
\bibitem [{\citenamefont {Wang}\ \emph {et~al.}(2015)\citenamefont {Wang},
  \citenamefont {Gresch}, \citenamefont {Soluyanov}, \citenamefont {Xie},
  \citenamefont {Dai}, \citenamefont {Troyer}, \citenamefont {Cava},\ and\
  \citenamefont {Bernevig}}]{Wang2015s}%
  \BibitemOpen
  \bibfield  {author} {\bibinfo {author} {\bibfnamefont {Z.}~\bibnamefont
  {Wang}}, \bibinfo {author} {\bibfnamefont {D.}~\bibnamefont {Gresch}},
  \bibinfo {author} {\bibfnamefont {A.~A.}\ \bibnamefont {Soluyanov}}, \bibinfo
  {author} {\bibfnamefont {W.}~\bibnamefont {Xie}}, \bibinfo {author}
  {\bibfnamefont {X.}~\bibnamefont {Dai}}, \bibinfo {author} {\bibfnamefont
  {M.}~\bibnamefont {Troyer}}, \bibinfo {author} {\bibfnamefont {R.~J.}\
  \bibnamefont {Cava}}, \ and\ \bibinfo {author} {\bibfnamefont {B.~A.}\
  \bibnamefont {Bernevig}},\ }\href
  {http://arxiv.org/abs/1511.07440} {\bibfield  {journal} {\bibinfo
  {journal} {arXiv:1511.07440}\ } (\bibinfo {year} {2015})}\BibitemShut
  {NoStop}%
\bibitem [{\citenamefont {Marzari}\ and\ \citenamefont
  {Vanderbilt}(1997)}]{Marzari1997}%
  \BibitemOpen
  \bibfield  {author} {\bibinfo {author} {\bibfnamefont {N.}~\bibnamefont
  {Marzari}}\ and\ \bibinfo {author} {\bibfnamefont {D.}~\bibnamefont
  {Vanderbilt}},\ }\href {\doibase 10.1103/PhysRevB.56.12847} {\bibfield
  {journal} {\bibinfo  {journal} {Phys. Rev. B}\ }\textbf {\bibinfo {volume}
  {56}},\ \bibinfo {pages} {12847} (\bibinfo {year} {1997})}\BibitemShut
  {NoStop}%
\bibitem [{\citenamefont {Souza}\ \emph {et~al.}(2001)\citenamefont {Souza},
  \citenamefont {Marzari},\ and\ \citenamefont {Vanderbilt}}]{Souza2001}%
  \BibitemOpen
  \bibfield  {author} {\bibinfo {author} {\bibfnamefont {I.}~\bibnamefont
  {Souza}}, \bibinfo {author} {\bibfnamefont {N.}~\bibnamefont {Marzari}}, \
  and\ \bibinfo {author} {\bibfnamefont {D.}~\bibnamefont {Vanderbilt}},\
  }\href {\doibase 10.1103/PhysRevB.65.035109} {\bibfield  {journal} {\bibinfo
  {journal} {Phys. Rev. B}\ }\textbf {\bibinfo {volume} {65}},\ \bibinfo
  {pages} {035109} (\bibinfo {year} {2001})}\BibitemShut {NoStop}%
\end{thebibliography}
\bibliographystyle{apsrev4-1}

\clearpage
\newpage
\onecolumngrid
\setcounter{equation}{0}
\setcounter{figure}{0}
\makeatletter

\setcounter{equation}{0}
\setcounter{figure}{0}
\setcounter{table}{0}
\setcounter{page}{1}
\setcounter{section}{0}
\makeatletter
\renewcommand{\theequation}{S\arabic{equation}}
\renewcommand{\thefigure}{S\arabic{figure}}
\renewcommand{\thetable}{S\arabic{table}}
\renewcommand{\thesection}{S\Roman{section}}
\renewcommand{\bibnumfmt}[1]{[S#1]}
\renewcommand{\citenumfont}[1]{S#1}

\section*{Supplemental Material for ``Universal signatures of Fermi arcs in quasiparticle interference on the surface of Weyl semimetals''}
\twocolumngrid

\section{Validity of JDOS / SSP description of the FTLDOS}

\begin{figure}[t]
 \centering
 \includegraphics[width=\columnwidth]{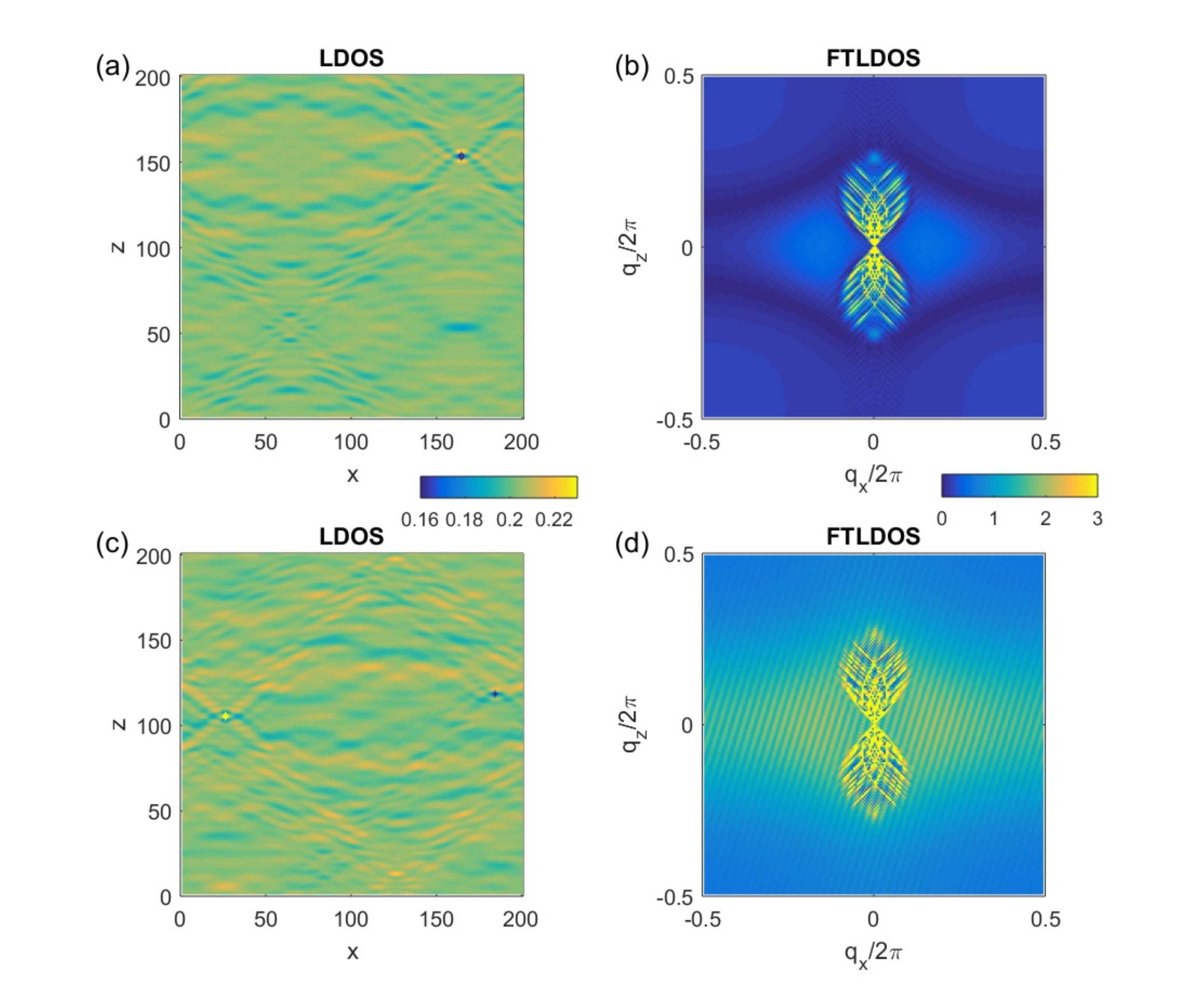}
 \caption{Two examples of simulated LDOS and FTLDOS (see text for definitions), both at $E = 0$, on a disordered $x$-$z$ surface of the simplest WSM [Eq.~(7) of main text]. The example shown in (a) and (b) corresponds to a surface that contains a single impurity, and the example shown in (c) and (d) corresponds to a surface that contains two impurities. The impurity centers can be located in the LDOS plots (a) and (c). The disorder strength (see text) is $W = 5$ in both examples. All other parameters are the same as in Fig. 2(a,b) in the main text.}
 \label{fig:ldos}
\end{figure}

In order to verify that the JDOS / SSP calculated in this paper indeed are qualitatively good approximations of the FTLDOS, we perform simulations of the LDOS on disordered surfaces of the simplest WSM corresponding to the hamiltonian of Eq.~(7) of the main text, and compare the FTLDOS obtained from the simulations with the JDOS shown in Fig. 2(b) of the main text. Explicitly, we add the following scalar potential term to the $x$-$z$ surface of the lattice model:
\begin{equation}
 V = \sum_{i=1}^{N_{\mathrm{imp}}} w_i \delta(\bm{r}-\bm{r}_i) \,,
\end{equation}
where $N_{\mathrm{imp}}$ is the number of surface impurities, $w_i$ is a random impurity potential uniformly distributed in the range $(-W/2,W/2)$ and $\bm{r}_i = (x_i,0,z_i)$ is the position of the $i$-th impurity placed randomly on the surface. The LDOS is then computed from the full Green's function by using the $T$-matrix method (cf. Eq. (2) in the main text)
\begin{align}
 \rho(\bm{r},E) = -\frac{1}{\pi} \mathrm{Tr} [ & G_{\bm{r},\bm{r}}(E)  \nonumber\\
 &+  \sum_{\bm{r}' ,\bm{r}''} G_{\bm{r},\bm{r}'}(E) \, T_{\bm{r}',\bm{r}''}(E) \, G_{\bm{r}'',\bm{r}}(E) ] \,,
\end{align}
where $G_{\bm{r},\bm{r}'}(E) = \int \mathrm{d}\bm{k} G(\bm{k},E) e^{\mathrm{i} \bm{k} (\bm{r}-\bm{r}')}$ is the unperturbed Green's function and $T_{\bm{r},\bm{r}'}(E) = V[1-G_{\bm{r},\bm{r}'}(E) V]^{-1}$ is the $T$-matrix associated with the impurity potential. We compute the LDOS for a finite but sufficiently large patch of the infinite surface and Fourier transform it to obtain the power spectrum of the FTLDOS, defined as
\begin{equation}
 \rho(\bm{q},E) = \left| \sum_{\bm{r}} \rho(\bm{r},E) e^{-\mathrm{i} \bm{q} \cdot \bm{r}} \right| \,.
\end{equation}
For the simplest WSM hamiltonian [Eq.~(7) of the main text], the LDOS and FTLDOS at $E=0$ are shown in Fig.~\ref{fig:ldos} for one or two impurities.

Even though additional characteristics are present in the FTLDOS compared to our corresponding JDOS / SSP results, our results show clearly that the salient features identified in the JDOS / SSP of WSMs in the main text --- i.e., the figure-eight pattern and pinch point --- are identical in the FTLDOS. Further, both JDOS and FTLDOS recover the same overall shape of the QPI pattern. We conclude that the universal characteristics of Fermi arcs we have identified in the QPI of WSMs are recovered regardless of the approximations used (if any) to obtain the QPI spectrum.

\section{Additional details on the modeling of surface scattering}

As discussed in the main text, in discovered and proposed WSM materials, crystal symmetries dictate the presence of several Weyl points in the bulk. This will generically lead to boundary Fermi surfaces with more than one Fermi arcs. When arcs are close to one another, inter-arc scattering patterns will overlap the intra-arc ones. Here we arrange two arcs in a way that exemplifies this point. The spectral function of the first arc is given by Eq.~(5) of the main text. The same equation defines the second arc, which is parametrized by $(\bm{k}_2,r_2,\gamma_2,\varphi_2)$; the total spectral function amounts to summing the spectral functions of the two arcs.

\begin{figure}[t]
 \centering
 \includegraphics[width=\columnwidth]{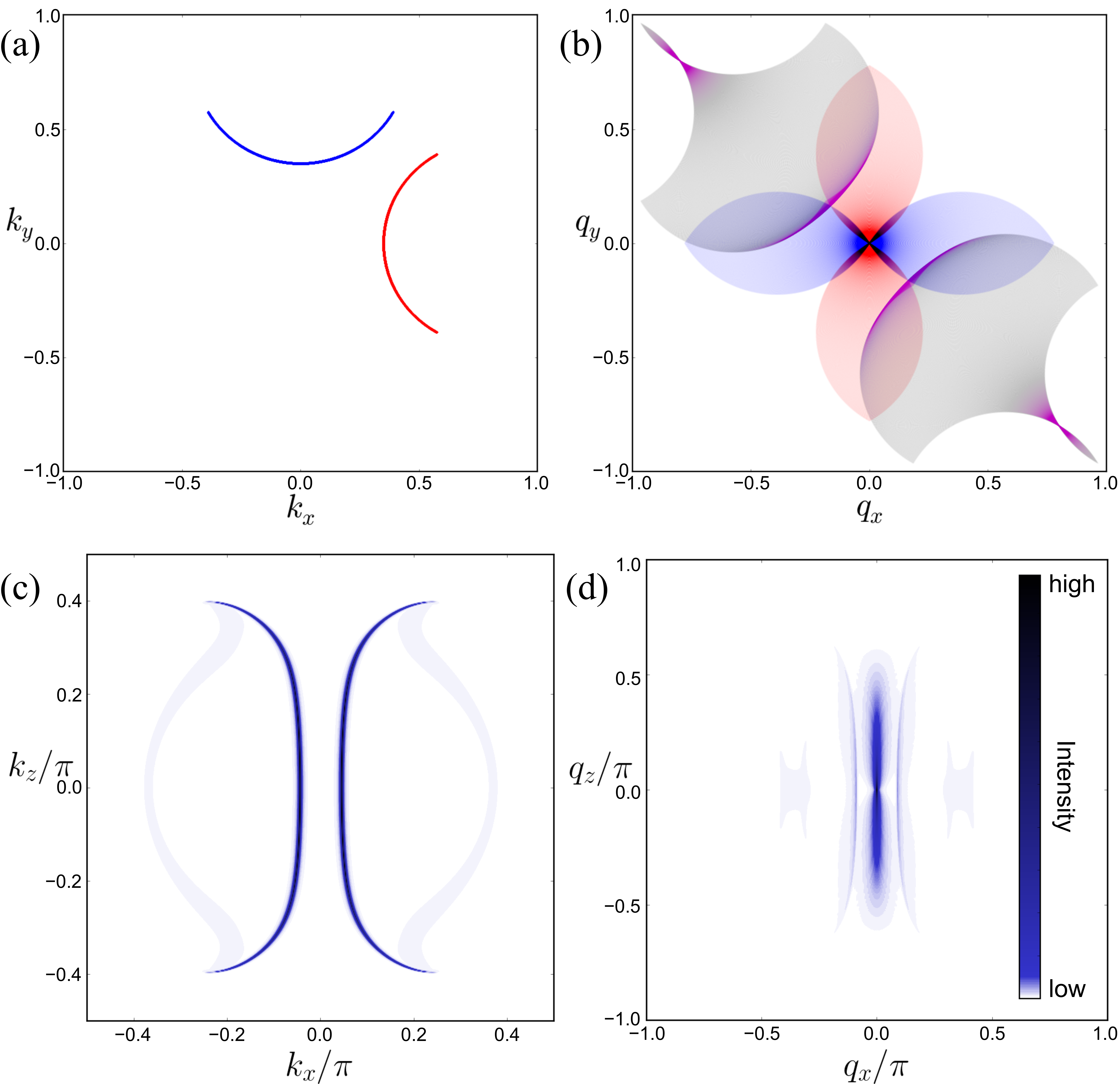}
 \caption{(a) Two Fermi arcs with $\gamma_1 = \gamma_2 = 2\pi/3$ and (b) the shape of the corresponding JDOS for a 90\degree arrangement; patterns in (b) are color-coded such that blue (red) pattern is generated by the autocorrelation of blue (red) contour in (a), whereas magenta signifies the cross-correlation between red and blue contours; (c) surface FS and (d) SSP for the model of Eq.~(8) with the parameters used for Figs.~2(c,d) of the main text at $E=-0.22$.}
 \label{fig:toysm}
\end{figure}

Overlapping contributions to the JDOS from more than one Fermi arcs can lead to finite spectral weight in the vicinity of $\bm{q}=0$, as seen in Fig.~\ref{fig:toysm}(b). Despite this fact, a superposition of characteristic ``figure-eights'' may still be discernible in $J_\nu$, since these contributions to the JDOS / SSP are additive. Furthermore, even though the cross-correlations of DOS corresponding to different arcs generally lack a distinct geometric identity, specific shapes can be traced back to the geometry of the underlying arcs. For example, one finds that partial nesting between different circular arcs results in pinch points in their cross-correlation at finite $\bm{q}$ [see Fig.~\ref{fig:toysm}(b)]. Such pinch points arise in the SSP of TaAs [see Fig.~4(b) of the main text and Fig.~\ref{fig:taassm}(b) here]. It is not always possible to unambiguously attribute the shapes of cross-correlations to a specific arc arrangement.

In both phenomenological and tight-binding modelings of Fermi arcs, we have assumed that the Fermi level is crossing the bulk bands exactly at the Weyl nodes, leading to a bulk FS consisting of isolated points. When the chemical potential is changed sufficiently, the projection of the bulk FS appears at the boundary. This situation is shown in Fig.~\ref{fig:toysm}(c) for the time-reversal symmetric tight-binding model used in the main text. For $E<0$, two oblong FS pockets appear at the ends of each arc. Upon deviating from $E=0$ further, the pockets increase in size and finally merge into a ribbon that connects the ends of each arc. The bulk contributions lead to new features and larger intensity around $\bm{q}=0$ in the SSP, but due to the fact that the Fermi arcs are the dominant surface FS feature, the figure-eight pattern and pinch point are recovered once again, in full analogy with our findings for MoTe$_2$.

In the tight-binding calculations, we obtain the Green's function $G$ at the surface of a semi-infinite system in the $y$ direction, with periodic boundary conditions in $x$ and $z$, using an iterative scheme for the evaluation of the surface transfer matrix~\cite{Sancho1984s,Sancho1985s}. We then evaluate the surface spectral function $A_0(k_x,k_z,E) = - \mathrm{Im \ Tr \ } G(k_x,k_z,E) / \pi$, as well as the spin-resolved spectral function $A_s(k_x,k_z,E) = (i/2\pi)\text{Tr}_{\tau} [G(k_x,k_z,E)-G(k_x,k_z,E)^\dag]$, and calculate their autocorrelations to obtain $J_0$ and $J_s$. In the band basis, $G$ is a diagonal matrix with entries $G_{bb}(\bm{k},E) = [E - \varepsilon_{b,\bm{k}} + \mathrm{i}\eta]^{-1}$, where $b$ is the band index and $\varepsilon_{b,\bm{k}}$ the band dispersion. When $\eta\rightarrow0$, the spectral functions become sums of delta functions. In this work, we assume a small but finite $\eta$, which broadens the delta functions into Lorentzians of half-width at half-maximum equal to $\eta$. This is done to emulate the broadening of the bands in real materials, as well as in our DFT calculations. Due to this broadening, there is a finite density in the vicinity of the pinch point at the origin [see Figs.~2(b,d)], caused by the finite width of the FS contour line. In particular, a finite lifetime broadening $\eta=0.005$ has been included in the Green's functions that yield Figs.~2(a,c) in the main text.

\section{Evaluation of the QPI pattern for TaAs}

\begin{figure}[t]
 \centering
 \includegraphics[width=\columnwidth]{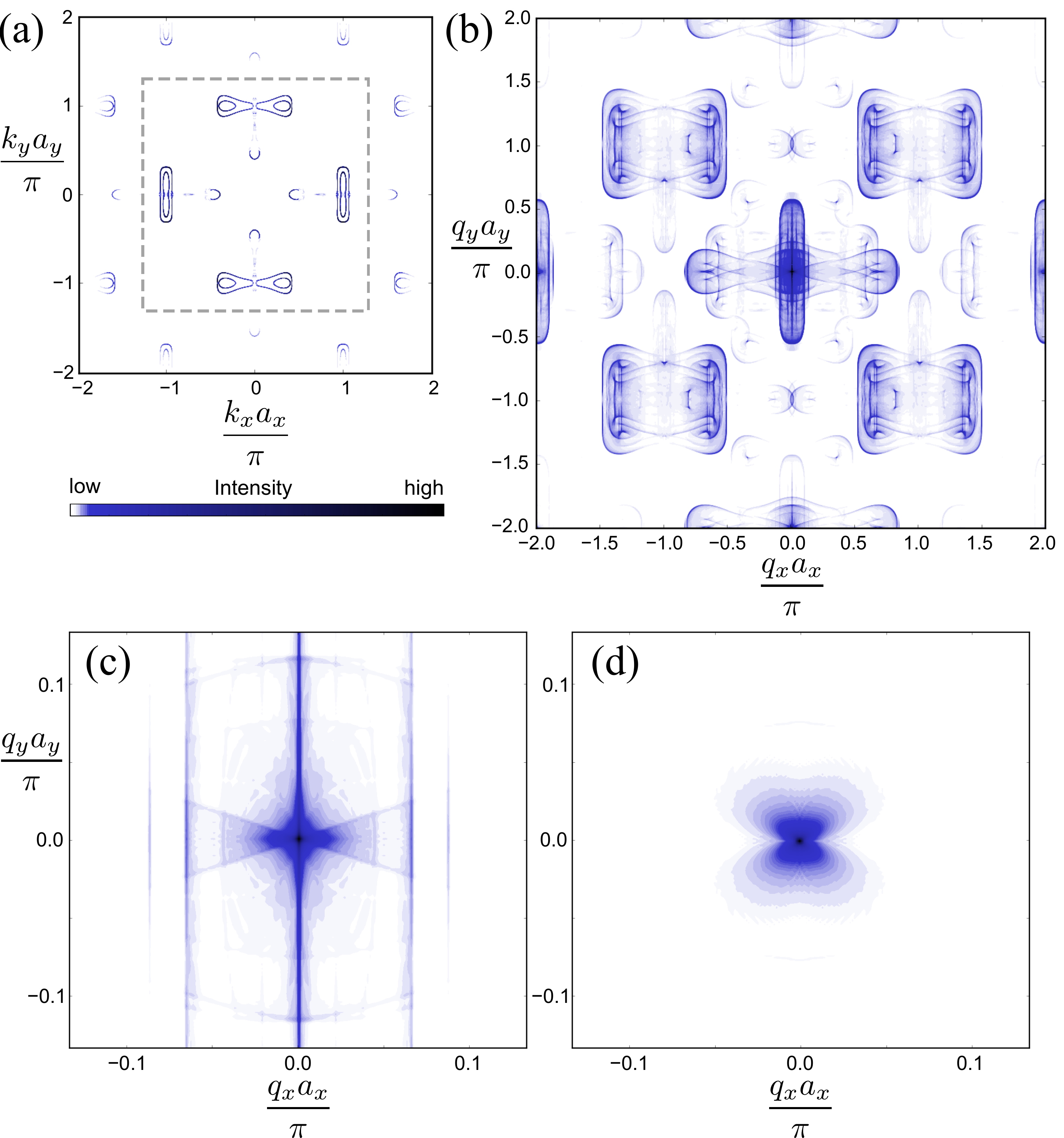}
 \caption{(a) FS at the (001) As-terminated surface of TaAs at $E=0.12$~eV, multiplied by the envelope function $F(\bm{k})$ defined in the text; (b) SSP using the full FS shown in (a); (c) SSP close to $\bm{q}=0$ for the area of the surface FS delimited by a dashed line in (a); (d) SSP in (c) minus SSPs centered at $(\pm2\pi,0)$ and $(0,\pm2\pi)$ --- compare to Fig.~4(h) of the main text.}
 \label{fig:taassm}
\end{figure}

The surface FS of TaAs contains segments that cross the boundary of the first BZ. Truncating the FS at the first BZ boundary would lead to false open contours, which would be misidentified as Fermi arcs in the QPI pattern. Such a truncation would furthermore be unnatural, as there is no physical reason for a sharp cutoff of scattering at a certain momentum. In practice, the overall QPI pattern does not depend very sensitively on whether one chooses to delimit the reciprocal space or not. However, the universal indicators for Fermi arcs become ambiguous when contours are artificially cut.

The more physical way to evaluate the QPI is to assume a form factor that envelops the spectral function. The reason for this is the interstitial spatial content of the rather localized Wannier wavefunctions used to simulate the electronic structure of a material~\cite{DellAnna2004,Choubey2014}. This leads to a decay of the measured DOS as one moves away from the first BZ center. To obtain the results shown in Fig.~4 of the main text, we have used the explicit form $F(\bm{k}) = \exp( -|\bm{k}|^2/\xi^2 )$ with $\xi=\pi$. The spectral functions in Eq.~(3) of the main text are multiplied by $F$ before the integration.

The decay of the form factor means that the QPI at points $(\pm2\pi,0)$ and $(0,\pm2\pi)$ comes predominantly from FS features close to the first BZ boundary, as scattering from the zone center at these wavevectors is heavily suppressed. One can use this to better resolve the QPI contributions close to $\bm{q}=0$ coming from FS features close to the zone center. In Fig.~\ref{fig:taassm} we illustrate the extreme case where the FS is truncated outside a certain range (i.e., the form factor becomes a theta function) so that the bow-tie features are included in their entirety while the spoon features in the second BZ are discarded. Then, by subtracting the QPI at $(\pm2\pi,0)$ and $(0,\pm2\pi)$ from the center of the pattern, we recover the autocorrelations of the Fermi arcs closest to the BZ center. The same procedure has been carried out in Fig.~4(g) for the more realistic case were the form factor decays smoothly, as described above. This method of uncovering QPI features in the vicinity of $\bm{q}=0$ may also be useful for the analysis of experimental data.

\section{QPI of Fermi arcs in MoTe${}_2$ at $E=0$}

\begin{figure}[t]
 \centering
 \includegraphics[width=\columnwidth]{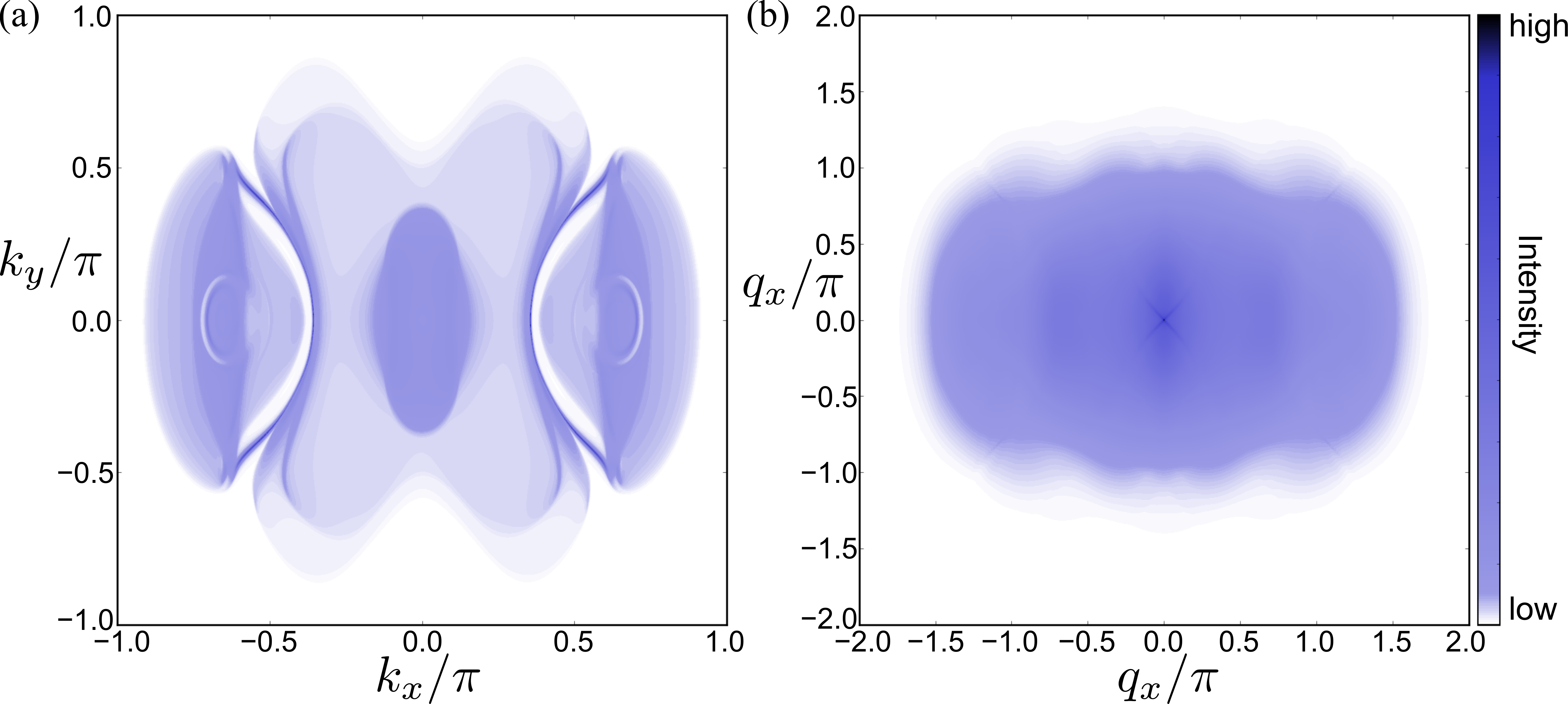}
 \caption{(a) Surface FS and (b) SSP for the (001) surface of MoTe${}_2$ at $E=0$~eV.}
 \label{fig:mote20}
\end{figure}

The signatures of Fermi arcs in the ab initio results presented in the main text do not have a sensitive dependence on the probing energy. To exemplify this, in Fig.~\ref{fig:mote20} we show the SSP for MoTe${}_2$ at $E=0$. Even though in this case the DOS intensities of bulk and boundary features are comparable, the ``X''-shaped scar identified in Fig.~3(b) of the main text is recovered, albeit not as pronouncedly as at $E=-0.05$~eV.

\section{Ab initio method}

Electronic structure calculations in this work were performed with density-functional theory (DFT) as implemented in the Vienna ab initio simulation package~\cite{Kresse1996}, and use the core-electron projector augmented wave basis sets~\cite{Blochl1994} with the generalized-gradient method~\cite{Perdew1996}. Spin-orbit coupling is included self-consistently. For the bulk calculation of MoTe$_2$, the cutoff energy for wave-function expansion is 350~eV, and the $\bm{k}$-point sampling grid is $16\times 10\times 4$. The experimental lattice parameters~\cite{Wang2015s} are used in calculation and convergence is checked with the above settings. The projected surface states are obtained from the surface Green function of the semi-infinite system~\cite{Wang2015s,Sancho1984s,Sancho1985s}. For this purpose, the maximally localized Wannier functions~\cite{Marzari1997,Souza2001} have been constructed from the first-principles calculations. 

For TaAs, due to the discrepancy of the similar Wannier-based surface calculations and unambiguous experimental observation, we preferred to perform a slab calculation of nine surface unit cells in thickness. The cutoff energy was chosen to be 300~eV. We used in-plane $\bm{k}$-point grids of size 12 $\times$ 12 for the charge self-consistent calculations, and size 1000 $\times$1000 for the FS calculations. The surface FSs were projected to the top unit cell of the As-terminated side.

The FSs are calculated in terms of the total spectral density
\begin{equation}
 \rho_0 = \mathrm{Tr} \bar A(\bm{k}) \,,
\end{equation}
as well as the spin density
\begin{equation}
 \rho_i = \mathrm{Tr}[ \sigma_i \bar A(\bm{k}) ] \,, \quad i = 1,2,3 \,,
\end{equation}
where $\bar A(\bm{k})$ is the spectral function matrix and $\sigma_{1,2,3}$ are the Pauli matrices for spin. For TaAs, all $\rho_0$ and $\rho_{1,2,3}$ are projected to the As-terminated surface layer by keeping $\bar A(\bm{k})$ only for the corresponding surface atoms, whereas MoTe${}_2$ terminates on Te. The JDOS is then calculated as
\begin{equation}
 J_0(\bm{q}) = \sum_{\bm{k}} \rho_0(\bm{k}) \rho_0(\bm{k} + \bm{q}) \,,
\end{equation}
while the SSP is given by
\begin{equation}
 J_s(\bm{q}) = \frac12 \sum_{i=0,1,2,3} \sum_{\bm{k}} \rho_i(\bm{k}) \rho_i(\bm{k} + \bm{q}) \,.
\end{equation}

\end{document}